\documentclass[lettersize,journal]{IEEEtran}
\usepackage{amsmath,amsfonts}
\usepackage{algorithmicx,algorithm}
\usepackage{array}
\usepackage[caption=false,font=normalsize,labelfont=sf,textfont=sf]{subfig}
\usepackage{textcomp}
\usepackage{stfloats}
\usepackage{url}
\usepackage{algpseudocode}
\usepackage{verbatim}
\usepackage{graphicx}
\usepackage{color}
\usepackage{amssymb}
\usepackage{bm}
\usepackage{epstopdf}
\usepackage{multirow}
\usepackage{cite}
\usepackage{fancyhdr}
\usepackage{booktabs}
\usepackage{hyperref}
\usepackage{float}
\usepackage{makecell}

\def\BibTeX{{\rm B\kern-.05em{\sc i\kern-.025em b}\kern-.08em
    T\kern-.1667em\lower.7ex\hbox{E}\kern-.125emX}}
\usepackage{balance}
\begin{document}
\title{{MVMS-RCN: A Dual-Domain Unified CT Reconstruction with Multi-sparse-view and Multi-scale Refinement-correction}}
\author{Xiaohong Fan, Ke Chen, Huaming Yi, Yin Yang, and Jianping Zhang
\thanks{{This work was supported by the Project of Scientific Research Fund of the Hunan Provincial Science and Technology Department (No. 2022RC3022, No.2023GK2029, No.2024ZL5017, No. 2024JJ1008), the science and technology
innovation Program of Hunan Province (No. 2024WZ9008), Program for Science and Technology Innovative Research Team in Higher Educational Institutions of Hunan Province of China.}}
\thanks{{Xiaohong Fan was with the School of Mathematics and Computational Science, Xiangtan University, and Hunan Key Laboratory for Computation and Simulation in Science and Engineering, Xiangtan 411105, China. He is now with the College of Mathematical Medicine, Zhejiang Normal University, Jinhua 321004, China (fanxiaohong@smail.xtu.edu.cn).}}
\thanks{Ke Chen was with the Centre for Mathematical Imaging Techniques, Department of Mathematical Sciences, The University of Liverpool, L6972L Liverpool, U.K. He is now with the Department of Mathematics and Statistics, University of Strathclyde, G1 1XH Glasgow, U.K. (k.chen@strath.ac.uk).}
\thanks{{Yin Yang and Huaming Yi are with the School of Mathematics and Computational Science, Xiangtan University, and Hunan National Applied Mathematics Center, Xiangtan 411105, China (yangyinxtu@xtu.edu.cn, yihuaming27@gmail.com).}}
\thanks{Jianping Zhang is with the School of Mathematics and Computational Science, Xiangtan University, and Key Laboratory of Intelligent Computing \& Information Processing of Ministry of Education, Xiangtan 411105, China. (jpzhang@xtu.edu.cn).}
\thanks{Corresponding authors: J. Zhang and Y. Yang.}
% \thanks{Our source codes are available at https://github.com/fanxiaohong/MVMS-RCN.}
}

\markboth{Journal of \LaTeX\ Class Files,~Vol.~18, No.~9, September~2020}%
{How to Use the IEEEtran \LaTeX \ Templates}

\maketitle

\begin{abstract}
X-ray Computed Tomography (CT) is one of the most important diagnostic imaging techniques in clinical applications. Sparse-view CT imaging reduces the number of projection views to a lower radiation dose and alleviates the potential risk of radiation exposure. Most existing deep learning (DL) and deep unfolding sparse-view CT reconstruction methods: 1) do not fully use the projection data; 2) do not always link their architecture designs to a mathematical theory; 3) do not flexibly deal with multi-sparse-view reconstruction assignments. This paper aims to use mathematical ideas and design optimal DL imaging algorithms for sparse-view CT reconstructions. We propose a novel dual-domain unified framework that offers a great deal of flexibility for multi-sparse-view CT reconstruction through a single model. This framework combines the theoretical advantages of model-based methods with the superior reconstruction performance of DL-based methods, resulting in the expected generalizability of DL. We propose a refinement module that utilizes unfolding projection domain to refine full-sparse-view projection errors, as well as an image domain correction module that distills multi-scale geometric error corrections to reconstruct sparse-view CT. This provides us with a new way to explore the potential of projection information and a new perspective on designing network architectures. The multi-scale geometric correction module is end-to-end learnable, and our method could function as a plug-and-play reconstruction technique, adaptable to various applications. Extensive experiments demonstrate that our framework is superior to other existing state-of-the-art methods. Our source codes are available at \href{https://github.com/fanxiaohong/MVMS-RCN}{https://github.com/fanxiaohong/MVMS-RCN.}
\end{abstract}


\section{Introduction} \label{sec:introduction}
X-ray Computed Tomography (CT), as a vital diagnostic imaging technique in clinical applications, poses a potential risk of radiation exposure. Filtered back projection (FBP) and algebraic reconstruction technique (ART) are classical full-view CT image reconstruction methods that are relatively sensitive to noise \cite{Natterer2001}. To reduce radiation dose and scanning time \cite{Pan2009,Abbas2013,Kim2015}, a sparse-view CT has been developed, which involves fewer projection sampling views. Unfortunately, this approach leads to even more ill-posed inverse problems \cite{Louis1986}, hindering the development of image reconstruction algorithms.

To deal with the highly ill-posed nature of the sparse-view CT imaging problem, the popular compressive sensing (CS) model of CT imaging uses regularization as
\begin{equation}
	\underset{\boldsymbol{x}}{\min} \left\{\frac{1}{2}\left\|\boldsymbol{P}_s \boldsymbol{x}-\boldsymbol{y}_{s}\right\|_{2}^{2}+\lambda\mathcal{W}(\boldsymbol{x})\right\},
	\label{eq2}
\end{equation}
where $\boldsymbol{y}_{s} \in \mathbb{R}^{q} ~(q=q_1 \times q_2)$ is a sparse-view projection observation, $q_1 $ and $q_2$ are the numbers of views and detector elements, respectively. $\boldsymbol{x} \in \mathbb{R}^{d}$ is an unknown image to be reconstructed, where $d=m_1\times m_2$, and $m_1$ and $m_2$ are the numbers of rows and columns of $x$, respectively. $\boldsymbol{P}_s \in \mathbb{R}^{q\times d}$ is a sparse-view projection transformation that models the CT imaging system. $\mathcal{W}(\boldsymbol{x})$ is a regularizer that incorporates image sparsity prior. $\lambda$ is a regularization parameter that should be determined based on the specific scenario.

During the past three decades, numerous new CT reconstruction models have emerged based on different priors of regularization. TV-based models employ piecewise constant/smoothness constraints to solve \eqref{eq2} for CT imaging \cite{LaRoque2008,Yang2010a}. Non-local/patch similarity is also used to reconstruct CT images \cite{Zeng2016}. Wavelets and framelets are employed to achieve high-quality CT image reconstruction \cite{Rantala2006,Dong2012}. Convolution sparse coding \cite{Bao2019,Lu2023}, low-rank \cite{Gao2011,Semerci2014,Kim2015}, and dictionary learning \cite{Zhao2012,Zhang2017b,Dong2018} are also employed for CT imaging. These approaches have the advantages of theoretical support and strong convergence, but they are computationally expensive and difficult to choose the image prior and optimal parameters.

Recently, deep learning (DL) has developed rapidly and achieved great success in sparse-view CT reconstruction \cite{Li2022a}. RED-CNN, a combination of an autoencoder, a deconvolution network, and shortcut connections, was proposed for low-dose CT imaging \cite{Chen2017}. FBPConvNet \cite{Jin2017} combines multiresolution decomposition and residual learning \cite{He2016} to reduce artifacts while preserving image details in sparse-view CT reconstruction. DD-Net \cite{Zhang2018c}, which combines a DenseNet \cite{Huang2017} with deconvolution, takes the FBP results as input for the CT reconstruction with a sparse view. Variants of U-Net \cite{Ronneberger2015}, such as dual-frame and tight-frame U-Nets, have been used to effectively recover high-frequency edges in CT with a sparse view \cite{Han2018}. To better regularize the low-dose CT denoising model, DU-GAN that uses U-Net-based discriminators in generative adversarial networks (GAN) was proposed to learn differences between the image and gradient domains \cite{Huang2022}. CTformer \cite{Wang2023}, a more powerful token rearrangement, has been proposed for low-dose CT denoising to incorporate local contextual information and avoid convolution.

The aforementioned CT image reconstruction methods do not utilize the prior information from the known projection observation $\boldsymbol{y}_{s}$, which limits their reconstruction performance. Several joint dual-domain reconstruction models have been proposed for low-dose CT imaging \cite{Zhou2022,CheslereanBoghiu2023}. CD-Net, which combines a projection domain network, an analytical reconstruction operator, and an image domain network, is designed to restore accurate anatomical information from both the projection and the image domains \cite{Zhang2021a}. DRONE, which integrates data-driven priors with kernel awareness used for compressed sensing, has been shown to produce better CT reconstruction results \cite{Wu2021}. DL techniques can reduce computational costs and achieve successful CT reconstruction results compared to model-based methods. However, these DL-based methods, which are trained with large amounts of data, are known as an unknown black box with limited theoretical understanding.

There are various ways to formulate and design a convolutional neural network (CNN) model, taking inspiration from CT imaging systems and reconstructing algorithms. LEARN (Learned Experts' Assessment-Based Reconstruction Network) \cite{Chen2018} is a "fields of experts"-based iterative reconstruction scheme that trains some stages for sparse-projection CT imaging. PD-Net \cite{Adler2018} incorporates a forward operator (possibly nonlinear) into a CNN architecture by unrolling a proximal-based primal-dual optimization method, with proximal operators replaced by CNN modules. ISTA-Net and ISTA-Net+ \cite{Zhang2018} are based on the Iterative Shrinkage-Thresholding Algorithm (ISTA) \cite{Daubechies2004,Elad2006}, and learn the proximal mapping associated with the sparsity-inducing regularizer using nonlinear transforms. FISTA-Net \cite{Xiang2020} unfolds the Fast Iterative Shrinkage/Thresholding Algorithm (FISTA) \cite{Beck2009} into a deep network, consisting of multiple gradient descent, proximal mapping (a proximal operator network for nonlinear thresholding), and momentum modules in cascade. AMP-Net \cite{Zhang2021} is established by unfolding the iterative denoising process of the approximate message passing (AMP) algorithm \cite{Donoho2009} rather than learning regularization terms. Nest-DGIL \cite{Fan2023a} based on the second Nesterov proximal gradient optimization has a powerful learning ability for high-/low-frequency image features and can theoretically guarantee the reconstruction of more geometric texture details. {DIVA \cite{Dutta2024} unfolded a baseline adaptive denoising algorithm (DeQuIP\cite{Dutta2021,Dutta2022}), relying on the theory of quantum many-body physics. It can handle non-local image structures through the patch-interaction term and the quantum-based Hamiltonian operator. DPIR \cite{Zhang2022a} plugs a deep denoiser prior into a half quadratic splitting based iterative algorithm to solve various image restoration problems.} These networks combine the advantages of model prior-based and learning-based approaches, such as the interpretability and generality of model-based methods, and the high efficiency and parameter tuning-free advantages
of DL-based methods \cite{Metzler2017,Borgerding2017,Aggarwal2019,Yang2020,Zhang2021b,Fan2021,Xia2021,Fan2023}.

However, there are some limitations of these imaging methods. 1) the existing methods generally only correct the projection error in single projection view; 2) the derivation from mathematical theory to network design is not always natural enough for these existing deep unfolding methods. In PD-Net, the forward and pseudoinverse operators lack thorough analysis and explanation, being substituted by CNNs. This problem is similarly found in the CNN denoisers utilized in AMP-Net and DPIR, and in the inversion of the degradation operator in DIVA. These mathematical operations are replaced by CNN without a clear theoretical foundation; 3) there is often a contradiction between real multi-sparse-view scenarios and existing sparse-view CT imaging methods that requires re-training for each given scenario.

To overcome these drawbacks, we design a refinement-correction architecture that consists of multi-view projection refinement in the projection domain and multi-scale geometric correction in the image domain. Specifically, 1) we refine projection errors from both the projected and non-projected views in both sparse-view and full-view settings, instead of solely focusing on the error measure from the projected views, giving a new way to explore the potential of available projection information in the projection domain; 2) we transform the completion of the details in the image domain into correcting different scales of errors about the target image geometric errors under multi-scale and multi-resolution. {By aligning the detailed mathematical framework with the network architecture, our approach ensures that all mathematical operations are systematically integrated into the network design.} The derivation from mathematical theory to network design is more natural than the existing deep unfolding methods; 3) the proposed unified dual-domain framework can train/predict multi-sparse-view CT reconstruction tasks through a single model to avoid expensive training costs and ample storage space, while the existing sparse-view CT methods have to train an independent model for each sparse-view CT task. The main contributions of this work can be summarized as follows.
\begin{itemize}
	\item[(1)] We propose a novel dual-domain unified framework that provides significant flexibility for multi-sparse-view CT reconstruction through a single model. This framework integrates the theoretical advantages of model-based methods with the superior reconstruction performance of DL-based methods, resulting in the expected generalizability of DL.
	\item[(2)] We propose a multi-view projection refinement module $\mathcal{R}$ that utilizes the unfolding projection domain to refine the full-sparse-view projection errors to reconstruct the multi-sparse-view CT. This provides us with a new way to explore the potential of projection information.
	\item[(3)] We propose an multi-scale geometric correction module $\mathcal{D}$ that distills multi-scale geometric error corrections to reconstruct the multi-sparse-view CT. This provides us with a new perspective on designing network architectures that systematically integrate mathematical operations.
	
	\item[(4)] The multi-scale geometric correction module $\mathcal{D}$ is learnable end to end, and our method could serve as a plug-and-play reconstruction technique suitable for various applications. Extensive experiments demonstrate that our framework is superior to other existing state-of-the-art methods.
\end{itemize}

\section{Background}
Monochromatic energy CT imaging can be formulated as a linear inverse problem:
\begin{equation}
	\boldsymbol{y}_{f}=\boldsymbol{P}_{f}\boldsymbol{x},
	\label{eq1-1}
\end{equation}
where $\boldsymbol{P}_{f}$ is a full-view projection transformation modeling the CT imaging, $\boldsymbol{y}_{f}$ is a full-view sampling observation, and $\boldsymbol{x}$ is an unknown image. The artifact-free image $\boldsymbol{x}_{f}$ can be approximately reconstructed by FBP as $\boldsymbol{x}_{f}=\boldsymbol{P}_{f}^{T}\boldsymbol{y}_{f}$ using all projections, and $\boldsymbol{P}_{f}^T$ is an FBP operator with the "Ram-Lak" filter in full-view.

The purpose of sparse-view CT reconstruction is to infer $\boldsymbol{x}_{f}$ from sparse-view projection observation $\boldsymbol{y}_{s}$ and $\boldsymbol{P}_{s}$. To simulate the sparse-view geometries, the sparse sampling $\boldsymbol{y}_{s}$ is modeled as follows:
\begin{equation}
	\boldsymbol{y}_{s} = \boldsymbol{P}_s \boldsymbol{x},
	\label{eq1}
\end{equation}
where $\boldsymbol{P}_s$ is a low-level transformation with sparse sampling.

The AMP algorithm \cite{Donoho2009} interprets a traditional linear reconstruction technique as the sum of the original data and a noise component. Consequently, the solution of \eqref{eq1} is estimated by utilizing two-step iterations as follows:
\begin{align}
	\mathbf{\boldsymbol{z}}_{\ell-1} &=\boldsymbol{y}_{s}-\boldsymbol{P}_s \boldsymbol{x}_{\ell-1},  \label{eq3} \\
	\boldsymbol{x}_{\ell} &=\text{Prox}^{\lambda}_{\mathcal{W}}(\boldsymbol{P}_s^{\mathrm{T}} \boldsymbol{z}_{\ell-1}+\boldsymbol{x}_{\ell-1}),
	\label{eq4}
\end{align}
where $\ell$ denotes the iteration stage, $\boldsymbol{P}_s^T$ is a FBP operator with the "Ram-Lak" filter in sparse-view, and
\[\text{Prox}^{\lambda}_{\mathcal{W}}(\bm{z})=\underset{\boldsymbol{x}}{\arg\min} \left\{\frac{1}{2}\left\|\boldsymbol{x}-\boldsymbol{z}\right\|_{2}^{2}+\lambda\mathcal{W}(\boldsymbol{x})\right\}\]
is a proximal-point operator that represents a geometric prior of the unknown image $\bm{x}$.

Using \eqref{eq1}, \eqref{eq3} and \eqref{eq4}, the denoising perspective of the AMP algorithm \cite{Zhang2021} is reformulated as follows:
\begin{equation}
	\begin{aligned}
		\boldsymbol{P}_{s}^{T} \boldsymbol{z}_{\ell-1}+\boldsymbol{x}_{\ell-1}
		&=\boldsymbol{P}_{s}^{T} \boldsymbol{P}_s({\boldsymbol{x}^*}-\boldsymbol{x}_{\ell-1})+\boldsymbol{x}_{\ell-1} \\
		&={\boldsymbol{x}^*}+(\boldsymbol{P}_s^{{T}} \boldsymbol{P}_s-\boldsymbol{I})\left({\boldsymbol{x}^*}-\boldsymbol{x}_{\ell-1}\right),
	\end{aligned}
	\label{eq5}
\end{equation}
where $\boldsymbol{e} = (\boldsymbol{P}_s^{{T}} \boldsymbol{P}_s-\boldsymbol{I})\left({\boldsymbol{x}^*}-\boldsymbol{x}_{\ell-1}\right)$ is the noise term, $\boldsymbol{x}^*$ is the true solution of \eqref{eq1}, $\boldsymbol{I}$ is the identity matrix. CNN with input $\boldsymbol{x}_{\ell-1}$ are used to learn the residual $\left({\boldsymbol{x}^*}-\boldsymbol{x}_{\ell-1}\right)$ \cite{Zhang2021}.

% ²åÈëÍ¼1
\begin{figure}
	\centerline{\includegraphics[width=1\columnwidth]{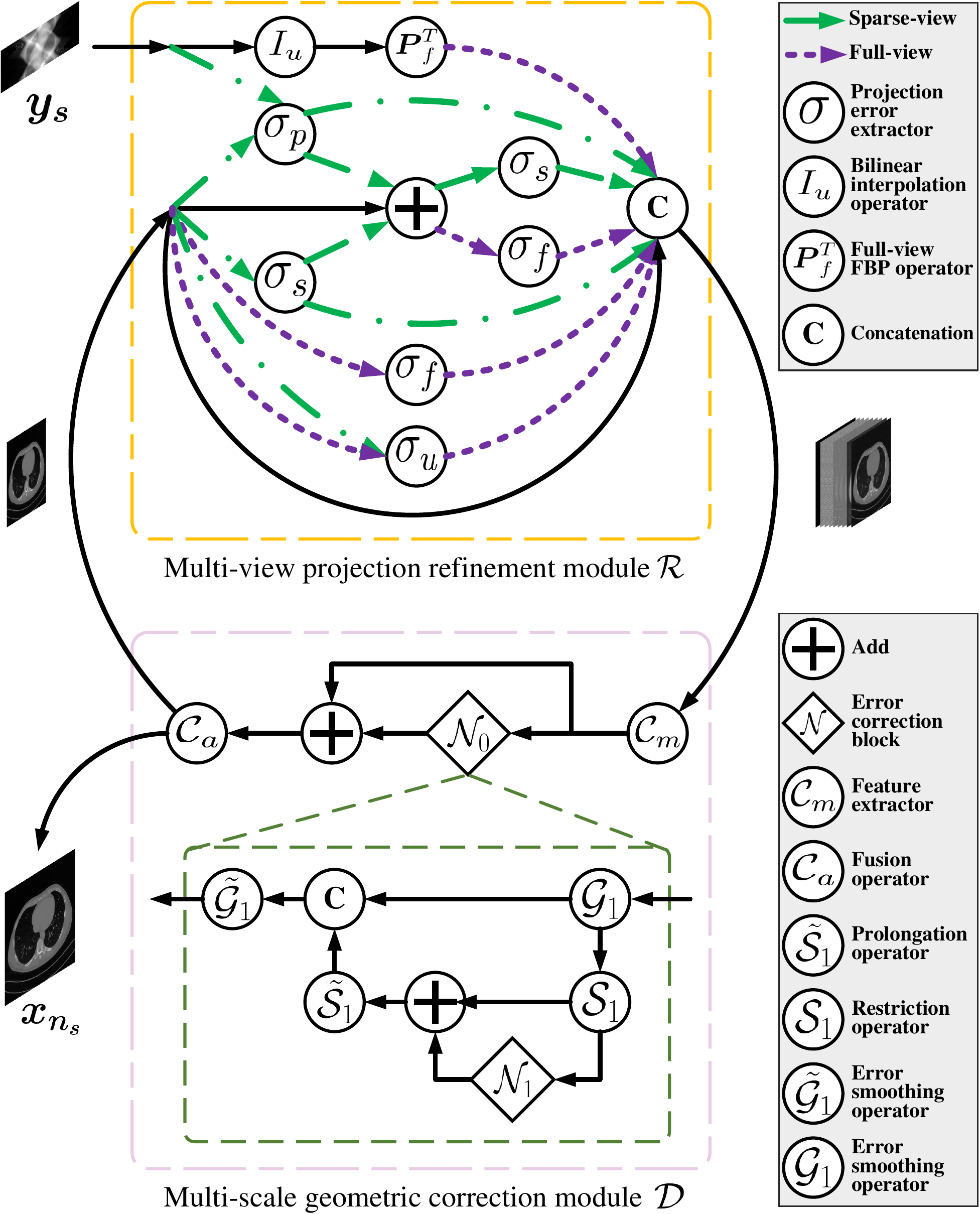}}
	\caption{{The overall architecture of the proposed unified dual-domain multi-sparse-view CT reconstruction framework (MVMS-RCN). It consists of multi-view projection refinement module $\mathcal{R}$ and multi-scale geometric correction module $\mathcal{D}$.}}
	\label{fig1}
\end{figure}

The challenge of solving \eqref{eq1} is its weak singularity, which can lead to numerical instability. To reduce the singularity, a naive approach is to compare the sparse-view reconstruction (SVR) error between $\boldsymbol{P}_s^T\boldsymbol{P}_s\boldsymbol{x}_{\ell-1}$ and  $\boldsymbol{P}_s^T\boldsymbol{P}_s\boldsymbol{x}^*$ by projecting the residual $\boldsymbol{z}_{\ell-1}$ on the projection domain, that is,
\[
\begin{split}
	\boldsymbol{e}_{\ell,s}&=\sigma_{p}(\boldsymbol{x}_{\ell-1},\boldsymbol{y}_{s},\boldsymbol{P}_s)\\&\equiv\boldsymbol{P}_s^T\mathbf{\boldsymbol{z}}_{\ell-1}
	=\boldsymbol{P}_s^T(\boldsymbol{y}_{s}-\boldsymbol{P}_s \boldsymbol{x}_{\ell-1})\\
	&=\boldsymbol{P}_s^T\boldsymbol{P}_s\boldsymbol{x}^*-\boldsymbol{P}_s^T\boldsymbol{P}_s\boldsymbol{x}_{\ell-1}=\boldsymbol{P}_s^T\boldsymbol{P}_s\boldsymbol{e}_{\ell-1,s},
\end{split}\]
{where $\sigma_{p}$ is the most commonly used operator for consistency checking of projection data.}

Most model-based DL methods for image reconstruction have been designed using the SVR error $\boldsymbol{e}_{\ell,s}$ for deep unfolding, which does not take full advantages of the projection data and the proximal-point operator $\text{Prox}^{\lambda}_{\mathcal{W}}$, leading to significant artifacts. In this work, we propose an interpretable dual-domain multi-scale CT reconstruction framework, MVMS-RCN, to create a novel approach to infer $\bm{x}_\ell$ from the current iteration $\bm{x}_{\ell-1}$. Our approach integrates the advantages of both model-based and DL-based techniques, enabling the correction of full-sparse-view projection errors in the projection domain and multi-scale geometric errors in the image domain, thereby facilitating the extraction of finer high-frequency details.

\section{Methodology}
The proposed framework shown in Fig. \ref{fig1} consists of a multi-view projection refinement module $\mathcal{R}$ and a multi-scale geometric correction module $\mathcal{D}$ to generate a high-quality approximation. The $\mathcal{R}$ module refines multi-view errors, which are categorized into four types: SVR error, two sparse-view projection (SVP) errors, two full-view projection (FVP) errors, and a cross-view reconstruction (CVR) error. This provides a novel way to extract high-quality information within the projection domain. Subsequently, the correction module $\mathcal{D}$, inspired by the multigrid scheme \cite{McCormick1987}, is used to correct multi-scale geometric errors of the target image.

\subsection{Multi-view Projection Refinement Module $\mathcal{R}$}
Unlike previous methods that utilize or learn directly from the SVR error \cite{Zhang2021a,Wu2021}, we explore several sparse-view and full-view projection errors to fully utilize the sampled projection data $\boldsymbol{y}_{s}$. We aim to merge full and sparse views to obtain multi-view corrections for the linear reconstruction error. This has the benefits of not only reducing the singularity of \eqref{eq1}, but also decomposing more texture details. It should be noted that the proposed error extraction method is based on the CT imaging projection mechanism rather than a black-box learning approach. The subsequent part provides details of the projection error extraction.

\textbf{Sparse-view projection (SVP).} Instead of explicitly solving the unknown $\bm{x}^*$ of \eqref{eq5}, which can be rewritten as
\begin{equation}
	\begin{aligned}
		{\boldsymbol{x}^*}=&\boldsymbol{P}_{s}^{T} (\boldsymbol{y}_{s}-\boldsymbol{P}_s \boldsymbol{x}_{\ell-1})+\boldsymbol{x}_{\ell-1}\\
		&-(\boldsymbol{I}-\boldsymbol{P}_s^{{T}} \boldsymbol{P}_s)\boldsymbol{x}_{\ell-1}+(\boldsymbol{I}-\boldsymbol{P}_s^{{T}} \boldsymbol{P}_s){\boldsymbol{x}^*},
	\end{aligned}
	\label{eq5a}
\end{equation}
we adopt the forward-backward splitting approach to approximately obtain the linear reconstruction solution $\boldsymbol{r}_{\ell}$ as follows
\begin{equation*}
	\begin{aligned}
		\boldsymbol{\hat{r}}_{\ell}=&\boldsymbol{x}_{\ell-1}+\boldsymbol{P}_s^{{T}} (\boldsymbol{y}_{s}-\boldsymbol{P}_s \boldsymbol{x}_{\ell-1})
		-(\boldsymbol{I}-\boldsymbol{P}_s^{{T}} \boldsymbol{P}_s)\boldsymbol{x}_{\ell-1},\\
		\boldsymbol{r}_{\ell}=&\boldsymbol{\hat{r}}_{\ell}+(\boldsymbol{I}-\boldsymbol{P}_s^{{T}} \boldsymbol{P}_s)\boldsymbol{\hat{r}}_{\ell}.
	\end{aligned} \label{eq5-1}
\end{equation*}

We estimate the linear reconstruction of the sparse-view projection for the multi-view projection refinement module $\mathcal{R}$ with respect to two SVP errors $\boldsymbol{e}_{\ell,d}$ and $\boldsymbol{e}_{\ell,j}$, where $\boldsymbol{e}_{\ell,d}$ and $\boldsymbol{e}_{\ell,j}$ are denoted respectively by
\begin{equation*}
	\begin{aligned}
		\boldsymbol{e}_{\ell,d}&=\sigma_{s}(\boldsymbol{x}_{\ell-1},\boldsymbol{P}_s)\equiv (\boldsymbol{I}-\boldsymbol{P}_s^{{T}} \boldsymbol{P}_s)\boldsymbol{x}_{\ell-1},\\
		\boldsymbol{e}_{\ell,j}&=\sigma_{s}(\boldsymbol{\hat{r}}_{\ell},\boldsymbol{P}_s)\equiv (\boldsymbol{I}-\boldsymbol{P}_s^{{T}} \boldsymbol{P}_s)\boldsymbol{\hat{r}}_{\ell},
	\end{aligned}
\end{equation*}
{where $\sigma_{s}$ is the sparse-view projection error extractor,} leading to two-step iterations defined by
\begin{equation}
	\begin{aligned}
		\boldsymbol{\hat{r}}_{\ell}&= \boldsymbol{x}_{\ell-1}+\boldsymbol{e}_{\ell,s}-\boldsymbol{e}_{\ell,d},\\
		\boldsymbol{r}_{\ell}&=\boldsymbol{\hat{r}}_{\ell} + \boldsymbol{e}_{\ell,j}.
	\end{aligned} \label{eq6}
\end{equation}

The total sparse-view error is combined using the extracted errors ($\boldsymbol{e}_{\ell,s}$, $\boldsymbol{e}_{\ell,d}$ and $\boldsymbol{e}_{\ell,j}$) in the projection domain. To facilitate the geometric errors correction in the image domain, we concatenate them together as follows
\[
\boldsymbol{e}_{\ell}^{\text{sparse}}=\textbf{Concat}\left(\boldsymbol{e}_{\ell,s},\boldsymbol{e}_{\ell,d},\boldsymbol{e}_{\ell,j}\right).
\]

\textbf{Full-view projection (FVP).} Next, we aim to extract more about the missing {projection data} under full-view \cite{Zhang2021a,Wu2021}, so that we can more accurately predict the linear reconstruction solution $\boldsymbol{r}_{\ell}$ from candidates with multiple views. To achieve this goal,
we refine two full view projection (FVP) errors $\boldsymbol{e}_{\ell,f}$ and $\boldsymbol{e}_{\ell,k}$ of the refinement module $\mathcal{R}$ in the projection domain, defined by the same operator $(\boldsymbol{I}-\boldsymbol{P}_{f}^{{T}} \boldsymbol{P}_{f})$ as follows
\[\begin{aligned}
	\boldsymbol{e}_{\ell,f}&=\sigma_{f}(\boldsymbol{x}_{\ell-1},\boldsymbol{P}_{f})\equiv (\boldsymbol{I}-\boldsymbol{P}_{f}^{{T}} \boldsymbol{P}_{f})\boldsymbol{x}_{\ell-1},\\
	\boldsymbol{e}_{\ell,k}&=\sigma_{f}(\boldsymbol{\hat{r}}_{\ell},\boldsymbol{P}_{f})\equiv (\boldsymbol{I}-\boldsymbol{P}_{f}^{{T}} \boldsymbol{P}_{f})\boldsymbol{\hat{r}}_{\ell},
\end{aligned}\]
where $\sigma_{f}$ is the full-view projection error extractor.

\textbf{Cross-view reconstruction (CVR).} The two-grid scheme \cite{Xu1992,Xu1996} has an intuitive appeal, as it suggests that a coarse grid solution can be used to improve the initial guess for a fine grid problem. This idea is consistent with the sparse-view CT reconstruction, suggesting that high-quality CT images can be reconstructed from sparse-view projection data. Therefore, we have developed a projection error refinement scheme that combines sparse view and full view, using the sparse view approximation
as an initial guess and one Newton iteration to correct the solution on the full view space. This scheme involves an upsampling interpolation and correction to produce a high-quality preliminary reconstruction.

To begin with, the sampling model $\boldsymbol{y}_{s} = \boldsymbol{P}_s \boldsymbol{x}$ is defined in the sparse view space (i.e. the coarse space). We then use FBP to reconstruct the CT image as follows:
\[\boldsymbol{x}_{u}= \boldsymbol{P}_{f}^T(\mathbb{I}_{u}\boldsymbol{y}_{s}),\]
where $\mathbb{I}_{u}$ is a bilinear interpolation operator and $\mathbb{I}_{u}\boldsymbol{y}_{s}$ is the
approximation on the full view space based on the initial guess $\boldsymbol{y}_{s}$ in the sparse view space. Subsequently, we project the current reconstruction $\boldsymbol{x_{\ell-1}}$ on sparse- and full-view spaces to estimate the interpolation projection error. This interpolation projection error corresponds to one Newton iteration to correct the solution $\mathbb{I}_{u}\boldsymbol{y}_{s}$ on the full-view space by using the two-grid scheme, and can be expressed as follows:
\[\boldsymbol{e}_{\ell,u}=\sigma_{u}(\boldsymbol{x}_{\ell-1},\boldsymbol{P}_{f},\boldsymbol{P}_s,\mathbb{I}_{u})\equiv \boldsymbol{P}_{f}^T(\mathbb{I}_{u}\boldsymbol{P}_s\boldsymbol{x}_{\ell-1}-\boldsymbol{P}_{f}\boldsymbol{x}_{\ell-1}),\]
where $\sigma_{u}$ is the cross-view projection error extractor.

The three projection errors ($\boldsymbol{e}_{\ell,u}$, $\boldsymbol{e}_{\ell,f}$, and $\boldsymbol{e}_{\ell,k}$)  that
have been extracted are given in full view in the projection domain. We merge them using a concatenation operator as follows:
\[\boldsymbol{e}_{\ell}^{\text{full}}=\textbf{Concat}\left(\boldsymbol{e}_{\ell,u},\boldsymbol{e}_{\ell,f},\boldsymbol{e}_{\ell,k}\right).\]

Recently, the addition of contextual information to traditional CNN architectures has been widely achieved through the use of channel concatenation. We fuse the full- and sparse-view projection errors using a concatenation operator and
refine the initial reconstruction $\boldsymbol{r}_\ell$, as follows:
\begin{equation}
	\begin{aligned}
		\boldsymbol{r}_\ell&=\mathcal{R}\left(\boldsymbol{x}_{\ell-1},\boldsymbol{y}_{s},\mathbb{I}_{u},\boldsymbol{P}_s,\boldsymbol{P}_{f}\right) \\
		&= \textbf{Concat}\left(\boldsymbol{x}_{\ell-1}, \boldsymbol{x}_{u},\boldsymbol{e}_{\ell}^{\text{full}},\boldsymbol{e}_{\ell}^{\text{sparse}}\right),
	\end{aligned}
	\label{eq9}
\end{equation}
where the initial input of stage 1 is set as $\boldsymbol{x}_{0}={\boldsymbol{P}_s^{T}}\boldsymbol{y}_{s}$.

It should be noted that we focus on extracting projection errors in the refinement module $\mathcal{R}$, which are based on the full-sparse-view projection mechanism in the projection domain. This provides us with a novel way to investigate the potential multi-view projection data in the projection domain. The architecture of the proposed multi-sparse-view projection error refinement module $\mathcal{R}$ is illustrated in Fig. \ref{fig1}.

\subsection{Multi-scale Geometric Correction Module $\mathcal{D}$}
Based on the preliminary reconstruction $\boldsymbol{x}_{\ell-1}$ and the full-sparse-view projection refinement correction, we obtain the intermediate reconstruction result $\boldsymbol{r}_{\ell}$ that is close to the target image $\boldsymbol{x}$. To further refine $\boldsymbol{r}_\ell$, we use an error correction formula $\mathcal{O}\left(\boldsymbol{x}\right)=\boldsymbol{x}-\boldsymbol{r}_{\ell}$ to estimate the difference between the target image $\boldsymbol{x}$ and $\boldsymbol{r}_{\ell}$. Since $\boldsymbol{x}$ is unknown, it is difficult to solve directly $\boldsymbol{x}$ from $(I-\mathcal{O})\left(\boldsymbol{x}\right)=\boldsymbol{r}_{\ell}$. If the nonlinear operator $\mathcal{O}$ satisfies the spectral constraint $\|\mathcal{O}\|<1$ \cite{Anselone1974}, we can use the Taylor expansion to split the nonlinear inverse operator $(I-\mathcal{O})^{-1}$ into different geometric levels for restoring $\boldsymbol{x}$. An approximation of the solution $\boldsymbol{x}$ can then be given as follows
\begin{equation}
	\begin{split}
		\boldsymbol{x}&=(I-\mathcal{O})^{-1}(\boldsymbol{r}_{\ell})=\left(I+\sum_{i=1}^{n} \mathcal{O}^{i}+\mathcal{E}(\mathcal{O}^{n})\right)\left(\boldsymbol{r}_{\ell}\right),
	\end{split}\label{eq8}
\end{equation}
where $\mathcal{E}\left(\mathcal{O}^{n}\right)(\cdot)$ computes the truncation remainder of the operator decomposition. $\mathcal{O}^{i}(\cdot)$ is used to extract different scale geometric errors to correct the intermediate reconstruction $\boldsymbol{r}_{\ell}$. $n$ is the multi-scale geometric error correction depth. In this study, we realize an optimal regularization prior (such as $\text{Prox}^{\lambda}_{\mathcal{W}}$ in \eqref{eq4}) by combining different correction operators $\mathcal{O}^{i}$,
thereby removing the necessity to determine a fixed $\lambda$.

Inspired by the classical multigrid scheme \cite{McCormick1987} and the design insights of multi-scale and multi-resolution networks, we propose a recursion multi-scale geometric error correction block $\mathcal{N}_i$ ($i=0,1,\dots,n-1$) (as shown in Fig. \ref{fig1}) embedded with convolution blocks and LeakyReLU layers for a more flexible implementation to approximate the geometric corrections $\sum_{k=i+1}^n\mathcal{O}^{k}(\cdot)$ and enhance the multi-scale errors extraction ability as follows:
\begin{equation}
	\begin{split}
		\mathcal{N}_{i}&=\tilde{\mathcal{G}}_{i+1}\left(I\cup\left[\tilde{\mathcal{S}}_{i+1}(I+\mathcal{N}_{i+1})\mathcal{S}_{i+1}\right]\right)\mathcal{G}_{i+1},
	\end{split}
	\label{block_N}
\end{equation}
where the convolution $\mathcal{S}_i$ and transpose convolution $\tilde{\mathcal{S}_i}$ ($2\times2$ kernel, stride 2 and $p$ channels, respectively) are employed to represent the restriction operator and the prolongation operator, respectively. $a\cup b=\textbf{Concat}(a,b)$ denotes the channel concatenation operation. The CNN blocks $\mathcal{G}_{i}$ and $\tilde{\mathcal{G}_{i}}$ are implemented as error smoothing operators on different error scales as follows:
% \begin{small}
	\[\begin{aligned}
		\mathcal{G}_{i}&= \operatorname{LeakyReLU}\left(
		\operatorname{Conv}\left[\operatorname{LeakyReLU}\left(\operatorname{Conv}\left[\cdot ; p, p\right]\right) ; p, p\right]\right),\\
		\tilde{\mathcal{G}_{i}}&= \operatorname{LeakyReLU}\left(
		\operatorname{Conv}\left[\operatorname{LeakyReLU}\left(\operatorname{Conv}\left[\cdot ; 2p, p\right]\right) ; p, p\right]\right),
	\end{aligned}\]
	% \end{small}
and $\operatorname{Conv}\left[\cdot ; p^{\mathrm{in}}, p^{\text {out }}\right]$ is the $3\times3$ convolution.

We can extend the proposed multi-scale geometric correction module $\mathcal{D}$ to any $n$ level in \eqref{eq8} as follows
\begin{equation}
	\begin{split}
		\boldsymbol{x}_{\ell}&=\mathcal{D}(\boldsymbol{r}_{\ell})=\mathcal{C}_{a}\left(I+\sum_{i=0}^{n-1} \mathcal{O}^{i+1}+\mathcal{E}(\mathcal{O}^{n})\right)\mathcal{C}_{m}(\boldsymbol{r}_{\ell})\\
		&\approx\mathcal{C}_{a}(I+\mathcal{N}_{0})\mathcal{C}_{m}(\boldsymbol{r}_{\ell})\\
		&=\mathcal{C}_{a}\left(I+\tilde{\mathcal{G}}_{1}\left(\left[I\cup\tilde{\mathcal{S}}_{1}(I+\mathcal{N}_1)\mathcal{S}_{1}\right]\mathcal{G}_{1}\right)\right)\mathcal{C}_{m}(\boldsymbol{r}_{\ell}),
	\end{split}
	\label{eq12}
\end{equation}
where $\mathcal{C}_{m}\left(\cdot\right)=\operatorname{LeakyReLU}\left(\operatorname{Conv}\left[\cdot; 1, p\right]\right)$ is a feature extractor. $\mathcal{C}_{a}\left(\cdot\right)=\operatorname{LeakyReLU}\left(\operatorname{Conv}\left[\cdot ; p, 1\right]\right)$ is a feature fusion operator. $\operatorname{Conv}\left[\cdot ; p^{\mathrm{in}}, p^{\text {out }}\right]$ is the $3\times3$ convolution. We set $\mathcal{N}_n=\mathcal{G}_{n+1}$ in this study. Many deep unfolding networks that take a single channel image as input have been shown to significantly hinder information transmission and lose many image details \cite{Song2021}. So we first transform the intermediate reconstruction $\boldsymbol{r}_{\ell}$ into features with $p$ channels through $\mathcal{C}_{m}\left(\cdot\right)$. Then we recover the corrected high-throughput information to a reconstructed CT image by $\mathcal{C}_{a}\left(\cdot\right)$.

It is worth mentioning that the same shared network parameters of the multi-scale geometric correction module $\mathcal{D}$ are used at all stages to ensure universality and possess the potential to be applied to plug-and-play reconstruction \cite{Aggarwal2019,Xiang2020}. The architecture of $\mathcal{D}$ is illustrated in Fig. \ref{fig1}.

The Algorithm \ref{algorithm1} and Fig. \ref{fig1} summarize the overall MVMS-RCN framework. This network is designed to perform multi-sparse-view CT reconstruction with a single model.
\begin{algorithm}[t]
	\caption{The Proposed MVMS-RCN Framework.}
	\label{algorithm1}
	
	{\bf Input:} The projection operators $\boldsymbol{P}_{f}$ and $\boldsymbol{P}_s$, the stage number $n_s$, FBP operators $\boldsymbol{P}_{f}^{T}$ and $\boldsymbol{P}_s^{T}$, the multi-scale depth $n$, the bilinear interpolation $\mathbb{I}_{u}$, the training dataset  $\mathcal{D}=\left\{\left(\boldsymbol{y}_s^{i},\boldsymbol{x}_f^{i}\right)\right\}_{i=1}^{N_d}$.\\
	{\bf Initialize:} $\boldsymbol{x}_0=\boldsymbol{P}_s^T\boldsymbol{y}_s$, the learnable parameters  $\boldsymbol{\Theta}=\left\{\mathcal{C}_{m},\mathcal{C}_{a},\left\{\mathcal{G}_{i},\tilde{\mathcal{G}}_{i},\mathcal{S}_{i},\tilde{\mathcal{S}}_{i}\right\}_{ i=1}^{n},\mathcal{G}_{n+1}\right\}$. \\
	{\bf Inference:} % \hspace*{0.24in}
	\begin{algorithmic}[1]
		\For{$\ell=1,2,\cdots,n_s$}
		\State $\boldsymbol{r}_\ell= \textbf{Concat}\left(\boldsymbol{x}_{\ell-1}, \boldsymbol{x}_{u},\boldsymbol{e}_{\ell}^{\text{full}},\boldsymbol{e}_{\ell}^{\text{sparse}}\right)$; // \eqref{eq9}
		\State $\boldsymbol{x}_{\ell}=\mathcal{C}_{a}(I+\mathcal{N}_{0})\mathcal{C}_{m}(\boldsymbol{r}_{\ell})$; // \eqref{block_N} and \eqref{eq12}
		\EndFor
	\end{algorithmic}
	{\bf Training:}
	\begin{algorithmic}[1]
		\State {$\mathcal{L}_{\text{total}}(\boldsymbol{\Theta})=\mathcal{L}_{\text{pixel}}(\boldsymbol{\Theta})+\gamma\mathcal{L}_{\text{SSIM}}(\boldsymbol{\Theta});$  \eqref{eq13}}
	\end{algorithmic}
	{\bf Output:}
	$\mathcal{M}(\mathcal{D} ; \boldsymbol{\Theta})=\boldsymbol{x}_{n_s}.$
\end{algorithm}

\subsection{Loss Function}
Sparse-view projection data $\boldsymbol{y}_{s}$ and initialization $\boldsymbol{x}_{0}$ are used as inputs to obtain the final reconstructed output $\boldsymbol{x}_{n_s}$ through the proposed method. The
loss function is then used to find the target image $\boldsymbol{x}^*$ by minimizing the distance between $\boldsymbol{x}_{n_s}$ and the original image
without artifacts $\boldsymbol{x}_{f}$. Here, we employ a $\ell_1$ loss instead of a $\ell_2$ loss which is not sufficient to capture perceptually relevant components (e.g., high-frequency geometric details) \cite{KJiang2018},  to increase the original pixel loss. Furthermore, we use a Structural Similarity Index Measure (SSIM) loss to quantify differences in brightness, contrast, and structure between $\boldsymbol{x}_{n_s}$ and $\boldsymbol{x}_{f}$, to help train the network. The total loss is expressed as follows:
\begin{equation}
\mathcal{L}_{\text{total}}(\boldsymbol{\Theta})=\mathcal{L}_{\text{pixel}}(\boldsymbol{\Theta})+\gamma\mathcal{L}_{\text{SSIM}}(\boldsymbol{\Theta})
\label{eq13}
\end{equation}
\[\text{with}: \begin{cases}\mathcal{L}_{\text{pixel}}(\boldsymbol{\Theta})=\frac{1}{d} \left\|\boldsymbol{x}_{\scriptscriptstyle n_s}-\boldsymbol{x}_{f}\right\|_{1}\\
	\mathcal{L}_{\text{SSIM}}(\boldsymbol{\Theta})=1-{\text{SSIM}}\left(\boldsymbol{x}_{\scriptscriptstyle n_s},\boldsymbol{x}_{f}\right),
\end{cases}\]
where $n_s$ is the total stage number of the proposed method. We set the weight coefficient $\gamma=1$.

\subsection{Parameters and Initialization}
Two modules in every stage of the proposed framework strictly correspond to the multi-view projection refinement module \eqref{eq9} and the multi-scale geometric correction module \eqref{eq12}, respectively. Learnable parameters $\boldsymbol{\Theta}=\left\{\mathcal{C}_{m},\mathcal{C}_{a},\left\{\mathcal{G}_{i},\tilde{\mathcal{G}}_{i},\mathcal{S}_{i},\tilde{\mathcal{S}}_{i}\right\}_{ i=1}^{n},\mathcal{G}_{n+1}\right\}$ consist of feature extractor $\mathcal{C}_{m}\left(\cdot\right)$, fusion operator $\mathcal{C}_{a}\left(\cdot\right)$, error smoothing operators $\mathcal{G}_{i}$, $\tilde{\mathcal{G}}_{i}$ and $\mathcal{G}_{n+1}$, restriction operator $\mathcal{S}_{i}$ and prolongation operator $\tilde{\mathcal{S}}_{i}$. All these parameters are learned as network parameters by minimizing the total loss $\mathcal{L}_{\text{total}}(\boldsymbol{\Theta})$.

Similarly to the traditional model-based method, the proposed method also requires an initial input $\boldsymbol{x}_{0}={\boldsymbol{P}_s^{T}}\boldsymbol{y}_{s}$ from the under-sampled projection data $\boldsymbol{y}_{s}$ in Fig. \ref{fig1}.
The network is initialized with Kaiming Initialization \cite{He2015}. The model parameters $\{p, n, n_{s}\}$ are initialized as $\{32, 5, 7\}$ respectively.

\section{Experiments and Results}
In this section, we illustrate the advantages of our MVMS-RCN in comparing with other reconstruction methods through various experiments. Peak signal to noise ratio (PSNR) and SSIM are used to evaluate their performances.

\subsection{Implementation Details} \label{data_detail}
The dataset used to evaluate the performances is from "Low Dose CT Image and Projection Data" \cite{Moen2020}.
{We chose the first ten-patient dataset of noncontrast chest CT scans to evaluate the reconstruction performances of the compared methods, which contains 3324 full-dose CT images of 1.5 mm thickness.} Eight patients' data have 2621 slices of 512 $\times$ 512 resolutions for training, one patient's data has 340 slices of 512 $\times$ 512 resolutions for validation, and the remaining one has 363 slices of 512 $\times$ 512 resolutions for testing. To increase the number of samples in the training dataset, horizontal and vertical flipping are applied. Fan-beam projection and parallel-beam projection, which are commonly used for sparse-view CT reconstruction, are used to sample the projection data.

For fan-beam projection, we employ a two-dimensional fan-beam geometry with 1024 angles evenly distributed over a full 360-degree radius. The detector is a plane containing 1024 pixels that are contiguous and spaced 2 mm apart. Both the source and the detector are positioned 500 mm away from the rotation center. The projection data with $1024\times1024$ resolutions are generated by full-dose CT sampling using the \textbf{RadonFanbeam} operator in TorchRadon \cite{Ronchetti2020}. These data can be down-sampled to 16, 32, 64, 128 and 256 views to simulate sparse-view geometries. The original artifact-free images are reconstructed by FBP using all 1024 views (full view).

For parallel-beam projection, the \textbf{Radon} operator in TorchRadon \cite{Ronchetti2020} is used to generate projection data with $720$ views and $729$ detectors through full-dose CT sampling. Additionally, the original artifact-free images are reconstructed by FBP using all 720 views (full-view). Moreover, the projection data can be down-sampled to 30, 60, 90, 120 and 180 views to simulate sparse-view geometries.

We use Pytorch to implement the proposed method with a batch size of 1 for sparse-view CT reconstruction. We use the Adam optimizer \cite{Kingma2014} with a learning rate of $1\times10^{-4}$ to train MVMS-RCN for 250 epochs. The experiments are conducted on a workstation with an Intel Xeon Silver 4214 CPU and a Tesla V100-PCIE-32GB GPU.

\subsection{Intra-Method Evaluation}
We perform several groups of experiments to determine the optimal configuration of the proposed MVMS-RCN for sparse-view CT reconstruction. These experiments involve the number of stages $n_s$, the depth of multiple scales $n$ within each stage, different initial inputs, different loss weight, an ablation study, and the sharing of modules.

\subsubsection{Stage number $n_s$}
We assess the reconstruction performance of different stage numbers by varying the stage number $n_s$ from 5 to 9. Table \ref{table_stage_number} shows the average PSNR for the fan-beam projection dataset with different views when the proposed framework is implemented with different stage numbers. To balance the complexity of the framework and the reconstruction performance \cite{Xiang2020}, we therefore fix $n_s=7$ for all subsequent experiments.

\subsubsection{Multi-scale depth $n$}
We investigate the effect of the multi-scale geometric error correction depth $n$ on the reconstruction performance. We vary the depth $n$ from 2 to 6. Table \ref{table_stage_number} presents the average PSNR values for the reconstructed CT images at different depths. It can be observed that the reconstruction performance gradually increases as the depth
increases, up to $n=5$. After this point, the performance becomes stable. We selected a depth of $n=5$ for optimal configurations, as it provides a good balance between network complexity and reconstruction performance.

\begin{table}%[width=\textwidth,cols=4, pos=h]
	\centering
	\caption{Quantitative assessment with PSNR values of different stage number $n_s$ and multi-scale depth $n$ by MVMS-RCN.}
	\label{table_stage_number}
	\setlength{\tabcolsep}{2.mm}{
		\begin{tabular*}{\hsize}{@{}@{\extracolsep{\fill}}
				cccccc@{}}      %}{ccccccc ccccccc}
		\toprule[1.5pt]
		{Stage $n_s$}
		& 5 & 6 & 7 & 8 & 9   \\
		\midrule[0.8pt]
		Params & 293441&293441&293441&293441&293441\\
		16 views &  38.34&38.39&\textbf{38.98}&37.60&38.26  \\
	   32 views & 41.49&41.44&\textbf{41.72}&41.47&41.53\\
	   64 views &43.23&43.25&\textbf{43.28}&43.20&43.11 \\
	   128 views &\textbf{44.86}&44.84&44.85&44.83& 44.69\\
	   256 views &\textbf{47.42}&47.33&47.27&47.31&47.08\\
	   Avg.& 43.07&43.05& \textbf{43.22}&42.88& 42.93\\
		\midrule[0.8pt]
		{Depth $n$}& 2&3 & 4 & 5 & 6 \\
		\midrule[0.8pt]
		Params & 130049&184513&238977&293441&347905\\
		16 views & 38.24&39.01&\textbf{39.04}&38.98&38.78\\
		32 views & 41.58&41.66&	41.59&\textbf{41.72}&41.57\\
		64 views & 43.25&43.26&43.20&\textbf{43.28}&43.17	\\
		128 views &44.81&44.80&44.77&\textbf{44.85}&44.77	\\
		256 views & \textbf{47.27}&47.26&47.24&\textbf{47.27}&47.19	\\
		Avg.& 43.03&43.20&43.17&\textbf{43.22}&43.10\\
		\bottomrule[1.5pt]
\end{tabular*}}
\end{table}

%%%%%%%%%%%%%%%%%%%%%%%%%%%%%%%%%%%%%%%%%%%%%%%%%%%%%%%
\begin{table}%[width=\textwidth, pos=htbp]
	\centering
	\caption{Quantitative assessment with different initialization of MVMS-RCN.}
	\label{table_initial}
	\setlength{\tabcolsep}{2.mm}{
		\begin{tabular*}{\hsize}{@{}@{\extracolsep{\fill}}
				cccccccc@{}}      %}{ccccccc ccccccc}
		\toprule[1.5pt]
		\multirow{3}{*}{Initialization}& \multicolumn{5}{c}{Views} & \multirow{3}{*}{Avg.}  \\\cmidrule(lr){2-6}
		 &16 & 32 & 64 & 128 &256 &  \\
		\midrule[0.8pt]
		$\boldsymbol{x}_0=\boldsymbol{P}_s^{T}\boldsymbol{y}_{s}$   & 38.98&41.72&43.28&44.85&47.27&43.22\\
		$\boldsymbol{x}_0=0$  & 38.98&41.70&43.28&44.81&47.25&43.20\\
	\bottomrule[1.5pt]
\end{tabular*}}
\end{table}

\subsubsection{Different initial inputs}
In this part, we evaluate the influence of two initialization schemes, $\boldsymbol{x}_0=\boldsymbol{P}_s^{T}\boldsymbol{y}_{s}$ and $\boldsymbol{x}_0=0$, on the proposed MVMS-RCN (fan-beam projection with 32 views). Table \ref{table_initial} shows the quantitative assessment of the different initial inputs $\boldsymbol{x}_0$. The reconstruction performance of $\boldsymbol{x}_0=0$ is slightly less than that of $\boldsymbol{x}_0=\boldsymbol{P}_s^{T}\boldsymbol{y}_{s}$, which validates the suitability of the initialization technique we employ, similar to numerous existing methods. Nevertheless, both initialization schemes yield good reconstruction performances, due to the effective error extraction of multi-view projection refinement and multi-scale geometric correction.

%%%%%%%%%%%%%%%%%%%%%%%%%%%%%%%%%%%%%%%%%%%%%%%
\begin{table}%[h]
	\centering
	\caption{{Quantitative assessment with different loss weight coefficient $\gamma$ of MVMS-RCN.}}
	\label{table_loss_weight}
	\setlength{\tabcolsep}{2.mm}{
		\begin{tabular*}{\hsize}{@{}@{\extracolsep{\fill}}
				cccccccc@{}}      %}{ccccccc ccccccc}
		\toprule[1.5pt]
		\multirow{3}{*}{Loss weight}& \multicolumn{5}{c}{Views} & \multirow{3}{*}{Avg.}  \\\cmidrule(lr){2-6}
		&16 & 32 & 64 & 128 &256 &  \\
		\midrule[0.8pt]
		$\gamma=0$ &38.68&41.71&43.31	&44.88	&47.37	&43.19\\
		$\gamma=0.1$ & 38.80&	41.74&	43.29&	44.85&	47.31&43.20\\
		$\gamma=1$   & 38.98&41.72&43.28&44.85&47.27&43.22\\
		$\gamma=10$ &38.27&41.44&	43.24&	44.86&	47.36&43.03\\
		\bottomrule[1.5pt]
\end{tabular*}}
\end{table}
%%%%%%%%%%%%%%%%%%%%%%%%%%%%%%%%%%%%%%%%%%%%%%%
\begin{table}%[width=\textwidth,cols=4, pos=h]
	\centering
	\caption{Quantitative assessment with PSNR values of different combinations of projection errors by proposed MVMS-RCN.}
	\label{table_ablation}
	\setlength{\tabcolsep}{1.mm}{
		\begin{tabular*}{\hsize}{@{}@{\extracolsep{\fill}}
				ccccccccc@{}}      %}{ccccccc ccccccc}
		\toprule[1.5pt]
		{Variant} &(a) &(b) & (c)&(d)&(e)& (f)&  (g)\\ \midrule[0.8pt]
		$\boldsymbol{x}_{\ell-1}$& + &+ &+ &+& +& +& +\\
		$\boldsymbol{x}_{u}~\& ~\boldsymbol{e}_{\ell,u}$  & - &+ & + &+ & -& -& +\\
		$\boldsymbol{e}_{\ell,f}~\& ~\boldsymbol{e}_{\ell,k}$ &-& -& +&+&- &- & +\\
		$\boldsymbol{e}_{\ell,s}$  &-&-&- &+&+ &+ & +\\
		$\boldsymbol{e}_{\ell,d}~\& ~\boldsymbol{e}_{\ell,j}$&-&- &- &-&-& +& +\\
		\midrule[0.8pt]
		%Parameters &291425&292001&292577&292865&291713&292289&293441\\
		16 views & 32.46&34.85&35.32&36.72&36.48&38.01&\textbf{38.98}\\
		32 views & 35.80&38.28&38.72&40.07&39.64&41.41&\textbf{41.72}\\
		64 views & 38.62&40.70&41.24&42.94&42.08&42.91&\textbf{43.28}\\
		128 views &	41.26&42.96&43.28&\textbf{45.11}&43.97&44.32&44.85\\
		256 views &42.98&44.95&45.21&\textbf{47.75}&46.02&46.32&47.27\\
		Avg.& 38.22&40.35&40.75&42.52&41.64&42.59&\textbf{43.22}\\
		\bottomrule[1.5pt]
\end{tabular*}}
\end{table}

%%%%%%%%%%%%%%%%%%%%%%%%%%%%%%%%%%%%%%%%%%%%%%%
\subsubsection{{Different loss weight}}
The SSIM loss measures luminance, contrast, and structural similarity, which aligns more closely with human visual perception than the $\ell_1$ loss. Typically, the SSIM loss results in more detailed reconstructions. In this study, the proposed approach addresses multi-sparse-view CT reconstruction challenges using a single model, making it impractical to fine-tune the weight parameter $\gamma$ for each specific task. Various combinations of weights for these losses were tested, and the results are shown in Table \ref{table_loss_weight}. Our results suggest that setting $\gamma$ to 1 is optimal.

%%%%%%%%%%%%%%%%%%%%%%%%%%%%%%%%%%%%%%%%%%%%%%%
\subsubsection{Ablation study}
Next, we conduct a group of ablation studies to evaluate the effectiveness of the SVR error $(\boldsymbol{e}_{\ell,s})$, SVP errors $(\boldsymbol{e}_{\ell,d},\boldsymbol{e}_{\ell,j})$, FVP errors $(\boldsymbol{e}_{\ell,f},\boldsymbol{e}_{\ell,k})$ and CVR $(\boldsymbol{x}_{u},\boldsymbol{e}_{\ell,u})$ on the performance of the CT reconstruction with sparse view. Comparisons are shown in Table \ref{table_ablation}. Unlike variant (a) which only relies on the previous output $\boldsymbol{x}_{\ell-1}$, variant (b) can use CVR reconstruction and error $(\boldsymbol{x}_{u},\boldsymbol{e}_{\ell,u})$ to greatly enhance CT reconstruction performance. From the comparison between variants (a) and (e) or the comparison between variants (c) and (d), the effectiveness of the widely used SVR error is apparent. From the comparison between variants (e) and (f) or the comparison between variants (d) and (g), the proposed SVP errors can further enhance the CT reconstruction performance. The comparison of variants (b) and (c) reveals that the proposed FVP errors can effectively improve CT reconstruction performance. From the above observations, we find that the proposed full-sparse-view projection error refinement techniques can effectively extract the potential projection information.

%%%%%%%%%%%%%%%%%%%%%%%%%%%%%%%%%%%%%%%%%%%%%%%%%%%
\begin{table*}%[htbp]
	\centering
	\caption{Quantitative assessment with different shared settings of MVMS-RCN.% Bold font in the table is the best of the variants.
	}
	\label{table_shared}
	\setlength{\tabcolsep}{2.mm}{
		\begin{tabular*}{\hsize}{@{}@{\extracolsep{\fill}}
				cccccccccc@{}}      %}{ccccccc ccccccc}
		\toprule[1.5pt]
		Variant &Shared setting& Params &16 views & 32 views & 64 views & 128 views &256 views & Avg.\\
		
		\midrule[0.8pt]
		(a)  &Shared $\mathcal{D}$ (default)  &293441 &\textbf{38.98}&\textbf{41.72}&\textbf{43.28}&\textbf{44.85}&47.27&\textbf{43.22}\\
		(b)  &Unshared &2054087&38.14&41.41&43.21&44.84&\textbf{47.28}&42.97\\
		\bottomrule[1.5pt]
\end{tabular*}}
\end{table*}

\begin{table*}%[width=\textwidth,cols=4, pos=h]
	\centering
	\caption{Quantitative sparse-view CT reconstruction performance comparisons of different methods. The best and second best results are highlighted in Bold font and underlined ones, respectively.}
	\label{table_result_CT}
	\setlength{\tabcolsep}{1.mm}{
		\begin{tabular*}{\hsize}{@{}@{\extracolsep{\fill}}clccccccccccc@{}}      %}{ccccccc ccccccc}
		\toprule[1.5pt]
		\multirow{3}{*}{{\shortstack{Projection\\Mode}}}
		&\multirow{3}{*}{{Method}}
		&\multicolumn{2}{c}{16 views ($\times$64)} & \multicolumn{2}{c}{32 views ($\times$32)} & \multicolumn{2}{c}{64 views ($\times$16)} & \multicolumn{2}{c}{128 views ($\times$8)} & \multicolumn{2}{c}{256 views ($\times$4)} & Time (s)\\ \cmidrule(lr){3-4} \cmidrule(lr){5-6} \cmidrule(lr){7-8}  \cmidrule(lr){9-10}  \cmidrule(lr){11-12}
		&&  PSNR  & SSIM  & PSNR & SSIM& PSNR & SSIM& PSNR & SSIM & PSNR & SSIM & CPU/GPU\\
		\midrule[0.8pt]
		\multirow{15}{*}{\rotatebox{90}{Fan-beam projection}} &FBP  &19.29 &0.1397&23.05 &0.2470 &26.67&0.4025&31.11&0.6268&36.46&0.8449&---/0.0021 \\
		&{FISTA-TV} \cite{Beck2009a} &25.20 &0.7024 & 26.87&0.7670 &28.77&0.8268&30.85&0.8836&32.82&0.9454&{116.54/---} \\
		&{RED-CNN} \cite{Chen2017} &32.15 & 0.8567 &36.59 &0.9057 &39.95&0.9361&42.02&0.9565&44.54&0.9709&---/0.0040 \\
		&{FBPConvNet} \cite{Jin2017} &32.50 & 0.8785 & 37.07&0.9175&40.14&0.9428& 42.82& 0.9588&44.86&0.9722&---/0.0075 \\
		&{DU-GAN} \cite{Huang2022} & 29.08&0.7629 &33.52 &0.8437 &36.58&0.8777&39.10&0.9123&42.04&0.9501&---/0.0021 \\
		&Uformer \cite{Wang2022} &32.58 & 0.8701&37.98 &0.9232&40.54&0.9436&41.93&0.9571&44.51&0.9707&---/0.1232 \\
		& {DRUNet \cite{Zhang2022a}} &{32.17}&{0.8870}&{37.50}&{\underline{0.9289}}&{40.82}&{0.9480}&{42.96}&{0.9612}&	{45.03}&{0.9736}
		&{---/0.0091}\\
		&{PD-Net}\cite{Adler2018} &\underline{34.49} & \underline{0.8877}&38.19 &0.9224 &40.35&0.9418&42.03&0.9551&43.73&0.9711&---/0.0243 \\
		&{ISTA-Net}\cite{Zhang2018} &31.65 & 0.8271&38.20&0.9177 & 41.94&0.9527&44.02&0.9671& \underline{46.41}&\underline{0.9803}&---/0.0401 \\
		&ISTA-Net+ \cite{Zhang2018} & 27.82 & 0.6434& 35.92&0.8921&42.23&0.9544 &44.17 &\underline{0.9679}&46.39& \underline{0.9803}&---/0.0520 \\
		&FISTA-Net \cite{Xiang2020} &28.87& 0.7062 &32.49 &0.8131  &35.82& 0.8692&38.10&0.8926&39.25&0.8875&---/0.1055 \\
		&AMP-Net-K \cite{Zhang2021} & --- & --- &26.39 &0.3767 &37.13& 0.8776&39.86&0.9265&41.87&0.9538&---/0.0304 \\
		&Nest-DGIL \cite{Fan2023a}& 29.25&0.8118&\underline{38.52}&{0.9269}&	\underline{42.69}&\underline{0.9570}&\underline{44.18}&0.9674&46.33&0.9800&---/0.0721\\
		& {DPIR \cite{Zhang2022a}} &{34.37}&{0.8869}&	{36.42}&{0.9147}&	{37.01}&{0.9279}&{37.22}&{0.9363}	
		 &{37.47}&{0.9443}& {---/4.1028}
		 \\
		&\textbf{MVMS-RCN} &\textbf{38.98} &\textbf{0.9363} &\textbf{41.72} &\textbf{0.9538} &\textbf{43.28}&\textbf{0.9636}&\textbf{44.85}&\textbf{0.9729}&\textbf{47.27}&\textbf{0.9844}&---/0.0836 \\
		\midrule[0.8pt]
		\multirow{3}{*}{{\shortstack{Projection\\Mode}}}
		&\multirow{3}{*}{{Method}}
		&\multicolumn{2}{c}{30 views ($\times$24)} & \multicolumn{2}{c}{60 views ($\times$12)} & \multicolumn{2}{c}{90 views ($\times$8)} & \multicolumn{2}{c}{120 views ($\times$6)} & \multicolumn{2}{c}{180 views ($\times$4)} & Time (s)\\ \cmidrule(lr){3-4} \cmidrule(lr){5-6} \cmidrule(lr){7-8}  \cmidrule(lr){9-10}  \cmidrule(lr){11-12}
		&&  PSNR & SSIM & PSNR & SSIM& PSNR & SSIM& PSNR & SSIM & PSNR & SSIM & CPU/GPU\\
		\midrule[0.8pt]
		\multirow{15}{*}{\rotatebox{90}{Parallel-beam projection}} &FBP  &23.98 &0.2979&28.69 & 0.5139&31.98&0.6796&34.67&0.7912&39.03&0.9051&---/0.0016 \\
		&{FISTA-TV} \cite{Beck2009a} &30.62 &0.8421& 35.17& 0.9045&38.13&0.9330&40.24&0.9509&43.56&0.9712&{36.461/---} \\
		&{RED-CNN} \cite{Chen2017} & 38.69& 0.9313 &42.01 & 0.9568&43.45&0.9656&44.44&0.9711&45.80&0.9783&---/0.0037 \\
		&{FBPConvNet} \cite{Jin2017} & 39.44 &0.9400 & 42.26 &0.9584 &43.50&0.9653&44.44&0.9708& 45.97&0.9788&---/0.0069 \\
		&{DU-GAN} \cite{Huang2022} &36.20 &0.8864 &38.70 & 0.9126&39.91& 0.9244&41.36&0.9434&43.20&0.9604&---/0.0020 \\
		&Uformer \cite{Wang2022} &40.08 &0.9437 &42.55 &0.9599 &43.67&0.9667&44.51&0.9714&45.89&0.9787&---/0.1243 \\
		& {DRUNet \cite{Zhang2022a}} &{40.37}&{0.9478}&{42.84}&{0.9617}&	{43.98}&{0.9683}&	{44.82}&{0.9728}&	{46.22}&{0.9798}&
		{---/0.0085}\\
		&{PD-Net}\cite{Adler2018} & 39.23&0.9368 &41.22 &0.9523 &42.26&0.9609&42.43&0.9632&43.85&0.9732&---/0.0247 \\
		&{ISTA-Net}\cite{Zhang2018} & \underline{41.06} &  \underline{0.9496}& 43.38&0.9644 &44.40&0.9704&45.17&0.9746&46.52&0.9811&---/0.0397 \\
		&ISTA-Net+ \cite{Zhang2018} & 41.00& 0.9490 &43.52 &0.9651 &{44.50}&0.9708&45.23&{0.9749}& 46.60&0.9814&---/0.0514 \\
		&FISTA-Net \cite{Xiang2020} &34.46 & 0.8540 &37.31 &0.9003 &39.28&0.9204&40.68&0.9418&41.65& 0.9515&---/0.1051 \\
		&AMP-Net-K \cite{Zhang2021} &37.67 &0.9063 &38.64 &0.9111 &43.55&0.9665& 42.72&0.9628&44.83& 0.9742&---/0.0315 \\
		&Nest-DGIL \cite{Fan2023a}&40.80&0.9468&\underline{43.77}&\underline{0.9664}&\underline{44.67}&\underline{0.9715}&\underline{45.38}&\underline{0.9754}&	\underline{46.72}&\underline{0.9817}&---/0.0998 \\
		& {DPIR\cite{Zhang2022a}}&{40.03}&{0.9462}&	{41.70}&{0.9592}&	{42.27}&{0.9643}&	{42.61}&{0.9676}&	{43.08}&{0.9724}&{---/4.0946} \\
		&\textbf{MVMS-RCN} & \textbf{42.38}&\textbf{0.9588} &\textbf{43.93} &\textbf{0.9676}&\textbf{44.80}&\textbf{0.9725}&\textbf{45.54}&\textbf{0.9764}&\textbf{46.91}&\textbf{0.9826}&---/0.0708 \\
		\bottomrule[1.5pt]
\end{tabular*}}
\end{table*}

\subsubsection{Module sharing configurations}
To demonstrate the superiority of the proposed multi-scale geometric correction module $\mathcal{D}$ with shared network parameter configurations among different stages, we conduct two variants of MVMS-RCN with different shared settings between stages. Table \ref{table_shared} lists the average PSNR comparisons with different shared settings of the proposed MVMS-RCN. The default shared variant (a) performs better than the unshared variant (b), indicating that the proposed multi-view
projection refinement module and the multi-scale geometric correction module are powerful enough to reduce the learning cost of the multi-scale geometric correction module $\mathcal{D}$ by sharing the learning parameters. We adopt the default shared variant (a) for the following experiments.

We attribute the superiority of our method to three factors. Firstly, our approach offers a dual-domain refinement-correction framework that can refine full-sparse-view projection errors and distill multi-scale geometric corrections to reconstruct sparse-view CT. Secondly, the projection error refinement from sparse-view and full-view can effectively extract the potential projection information. Lastly, a multi-scale geometric correction module inspired by the multigrid scheme can effectively correct the geometric error under multi-scale and multi-resolution.

% fig box
\begin{figure*}
	\centerline{\includegraphics[width=2\columnwidth]{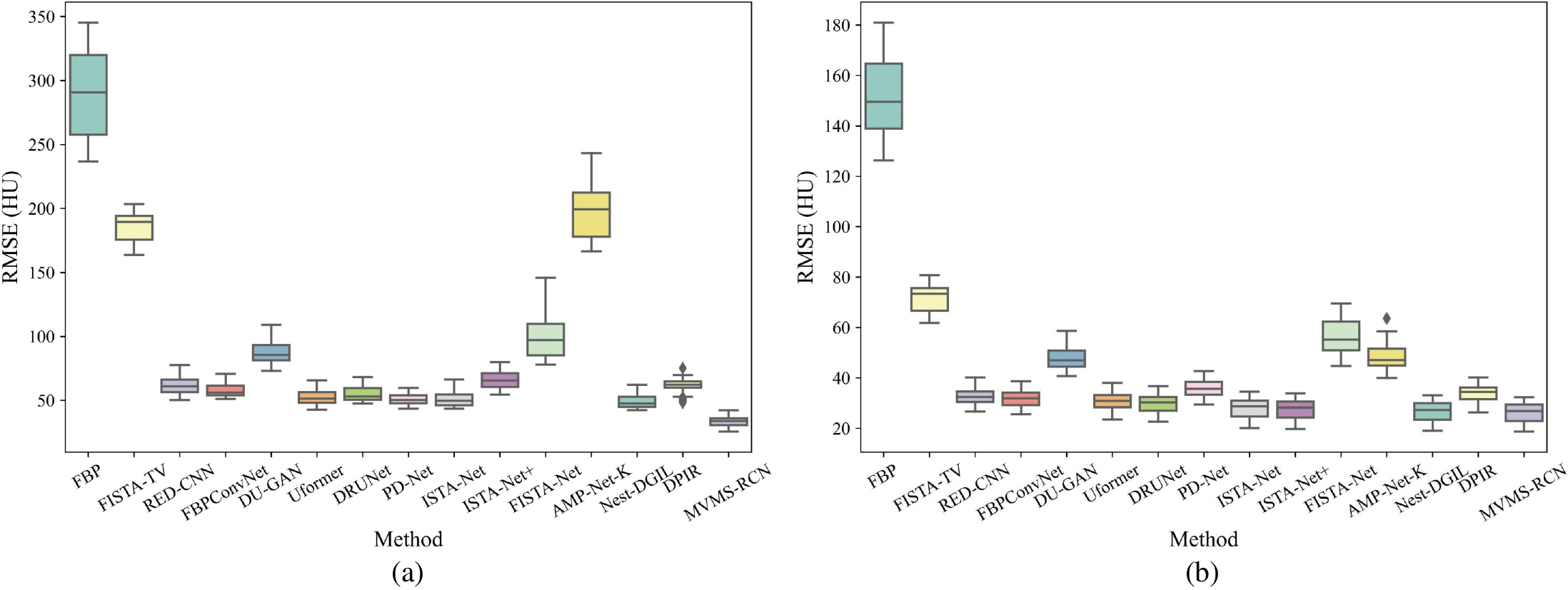}}
	\caption{{Boxplots of RMSE (HU) values of different methods for (a) fan-beam projection with 32 views and (b) parallel-beam projection with 60 views.}}
	\label{fig_box_rmse}
\end{figure*}

% fan-beam result comparison
\begin{figure*}
	\centerline{\includegraphics[width=1.97\columnwidth]{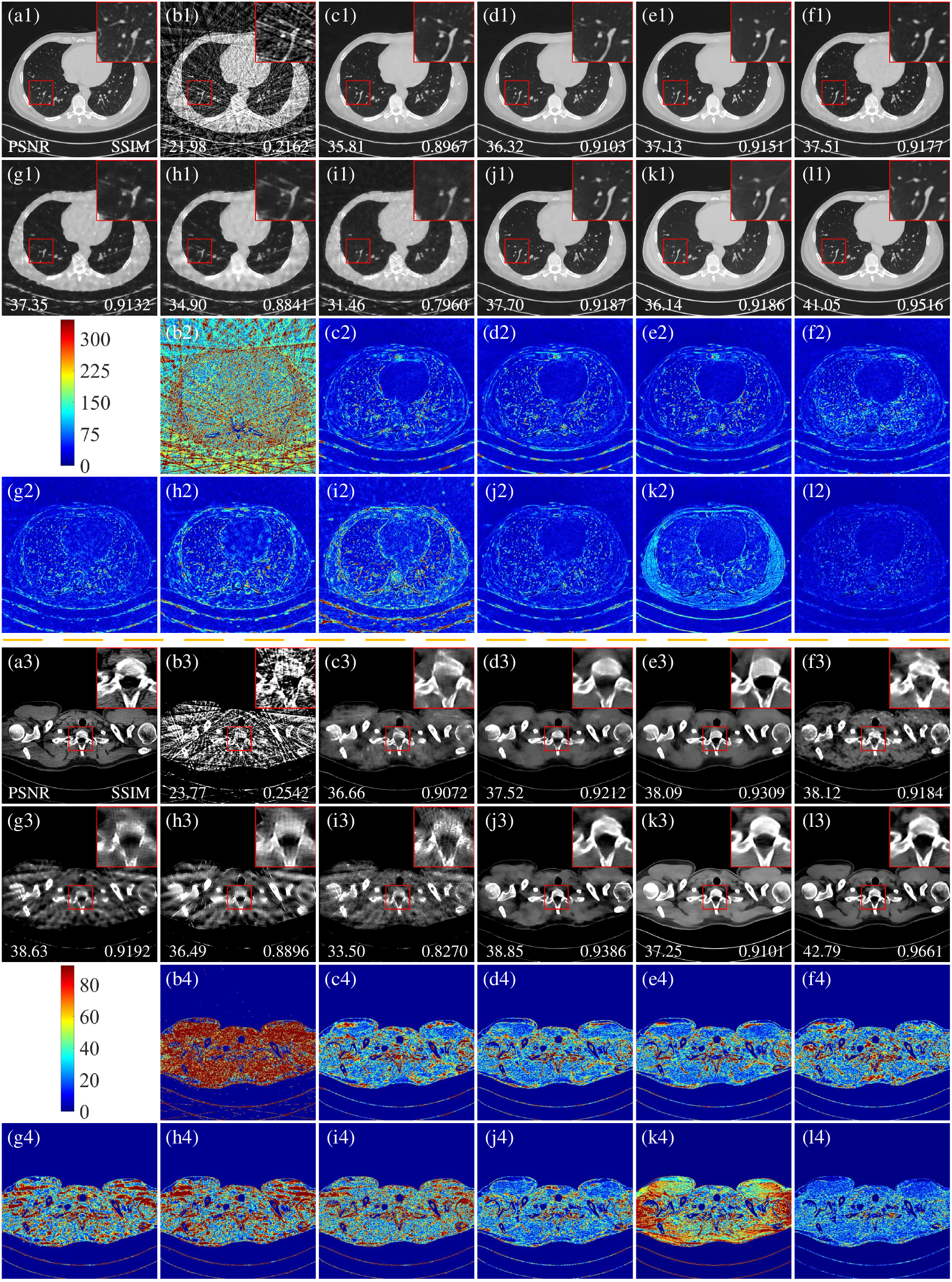}}
	\caption{{The axial reconstruction results ('1' and '3') and corresponding residuals ('2' and '4') from different methods for fan-beam projection with 32 views. (a1),(a3) reference image, (b1)-(b4) FBP, (c1)-(c4) RED-CNN, (d1)-(d4) FBPConvNet, (e1)-(e4) Uformer, (f1)-(f4) PD-Net, (g1)-(g4) ISTA-Net, (h1)-(h4) ISTA-Net+, (i1)-(i4) FISTA-Net, (j1)-(j4) Nest-DGIL, (k1)-(k4) DPIR, (l1)-(l4) MVMS-RCN. The display windows are [-1150, 350] HU and [-160, 240] HU, respectively.}}
	\label{fig_comparison_CT}
\end{figure*}

\begin{figure*}
	\centerline{\includegraphics[width=2\columnwidth]{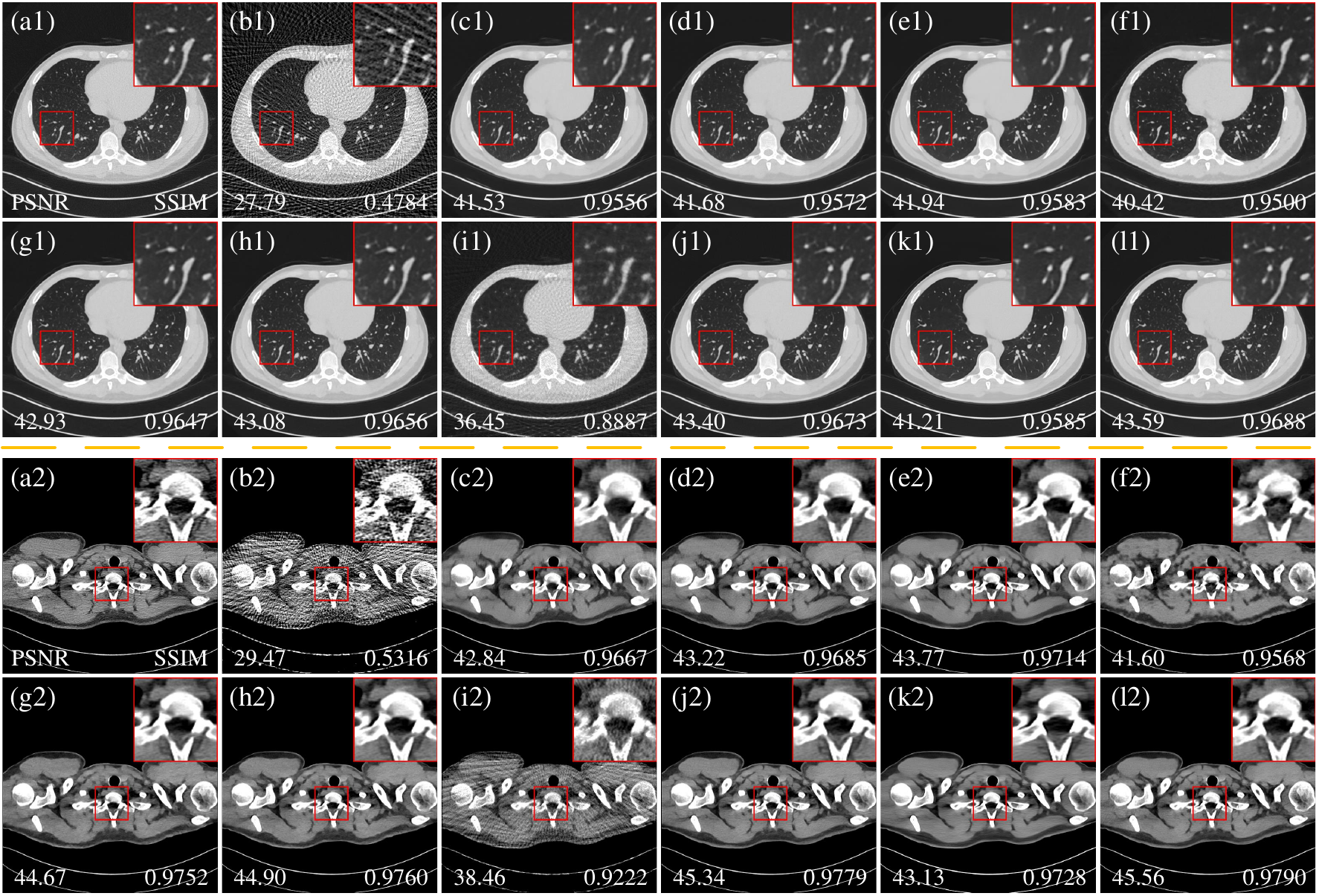}}
	\caption{{The axial reconstruction results from different methods for parallel-beam projection with 60 views. (a1)-(a2) reference image, (b1)-(b2) FBP, (c1)-(c2) RED-CNN, (d1)-(d2) FBPConvNet, (e1)-(e2) Uformer, (f1)-(f2) PD-Net, (g1)-(g2) ISTA-Net, (h1)-(h2) ISTA-Net+, (i1)-(i2) FISTA-Net, (j1)-(j2) Nest-DGIL, (k1)-(k2) DPIR, (l1)-(l2) MVMS-RCN. The display windows are [-1150, 350] HU and [-160, 240] HU, respectively.}}
	\label{fig_comparison_CT_parallel}
\end{figure*}

\subsection{Comparison to the State-of-the-art Methods}\label{section_CM}
{To further demonstrate the superiority of our approach, we conduct a group of studies to evaluate the reconstruction performance of fourteen compared methods on sparse-view CT. We compare our MVMS-RCN with traditional approaches such as FBP and FISTA-TV \cite{Beck2009a}), as well as DL techniques such as RED-CNN \cite{Chen2017}, FBPConvNet \cite{Jin2017}, DU-GAN \cite{Huang2022}, Uformer \cite{Wang2022} and DRUNet \cite{Zhang2022a}}. {Furthermore, we compare with deep unfolding methods, including PD-Net \cite{Adler2018}, ISTA-Net \cite{Zhang2018}, ISTA-Net+ \cite{Zhang2018}, FISTA-Net \cite{Xiang2020}, AMP-Net-K \cite{Zhang2021} and Nest-DGIL \cite{Fan2023a}, as well as the plug-and-play image restoration method DPIR \cite{Zhang2022a}.} PD-Net stands out as a tried-and-true approach for dual-domain reconstruction. Likewise, other deep unfolding methods also perform corrections in both the image domain and the projection domain. The FBP with the "Ram-Lak" filter, which was implemented with the Iradon transform \cite{Ronchetti2020}, is utilized as the initial step for the proposed methods and other compared methods. The maximum number of iterations for FISTA-TV is set to 100, and the regularization parameter is tuned to its optimal value. As reported in \cite{Xiang2020}, the number of stages for PD-Net, ISTA-Net, ISTA-Net+, FISTA-Net, AMP-Net-K, Nest-DGIL and our approach are all configured as $7$. A DRUNet model designed for sparse-view CT reconstruction is developed by changing the training data and replacing the noise level with the sparse-view down-sample scale. Meanwhile, the DPIR denoiser is re-trained using CT data. In DPIR, the data subproblem is tackled with the conjugate gradient (CG) method, which runs for five iterations. In addition, the outer iterations for DPIR are set to 40. Each of the compared methods is trained using an identical training dataset and employs the same techniques as described in their respective works.

\subsubsection{Quantitative evaluation}
Table \ref{table_result_CT} shows the average PSNR and SSIM values of the reconstruction results of the compared methods with different projection views of the downsampling. {Clearly, the bottom row of Table \ref{table_result_CT} shows the superior results.} {The DPIR denoiser trained on Gaussian noise cannot effectively remove the noise generated by sparse-view sampling.} All of the DL methods and deep unfolding methods have somewhat competitive reconstruction results for fan-beam and parallel-beam projection data; however, our methods clearly outperform the other compared methods in all reconstruction tasks from different down-sampling projections. In particular for fan-beam projection, the proposed MVMS-RCN have a 4.49 dB improvement with 16 views, a 3.20 dB improvement with 32 views, and an average 1.96 dB improvement on five sparse view CT reconstruction tasks. Additionally, our methods have an average 0.39 dB improvement for parallel-beam projection.

In Fig. \ref{fig_box_rmse}, we display the boxplots illustrating the Root Mean Square Error of Hounsfield Unit (RMSE (HU)) values for the fan-beam projection with 32 views and the parallel beam projection with 60 views to further highlight the superiority of our methods. Our methods are able to balance the trade-off between the use of prior knowledge from models and the powerful learning ability of DL, resulting in excellent generalizability. Furthermore, our methods outperform the compared methods in terms of reconstruction performance.

\begin{table*}%[width=\textwidth,cols=4, pos=h]
	\centering
	\caption{Quantitative reconstruction comparisons (PSNR) of untrained sparse-view projections. The best and second best results are highlighted in Bold font and underlined ones, respectively.}
	\label{table_result_untrained}
	\setlength{\tabcolsep}{1.mm}{
		\begin{tabular*}{\hsize}{@{}@{\extracolsep{\fill}}lcccccccc@{}}      %}{ccccccc ccccccc}
			\toprule[1.5pt]
			\multirow{3}{*}{{Method}}
			&\multicolumn{4}{c}{Fan-beam projection} &\multicolumn{4}{c}{Parallel-beam projection} \\ \cmidrule(lr){2-5}  \cmidrule(lr){6-9}
			&  24 views & 48 views &96 views & 192 views &  45 views & 75 views &105 views & 150 views\\
			\midrule[0.8pt]
			FBP  & 21.31&25.15&29.40&33.95& 26.62 &30.41&33.32&36.97 \\
			{FBPConvNet} &33.70/34.22&38.38/38.59&41.28/41.41&43.24/42.64&39.93/40.13&42.68/42.46&43.83/43.73&44.80/44.62\\
			{Uformer} &34.21/{35.57}&38.89/39.28&41.66/41.53&43.30/42.65&41.00/41.11&43.07/42.84&44.02/43.90&45.01/44.68\\
			{DRUNet} &{34.99} &	{39.25} &	{41.92} &	{43.50} &{41.39 }&	{43.36} &	{44.37} &	{45.34 } \\
			{PD-Net} & 32.85/28.99&35.30/28.53&35.95/30.79&33.52/33.88& 34.64/32.02&38.17/37.98&40.23/39.73&38.53/38.15\\
			{ISTA-Net} & 33.84/29.14&\underline{39.65}/31.23&42.37/39.80&44.44/44.06& 42.03/24.70&43.87/43.71&44.75/44.72&45.72/45.66\\
			ISTA-Net+ &  29.50/30.84&37.65/38.39&{42.91}/42.53&\underline{44.67}/44.37&\underline{42.07}/40.21&{43.99}/43.93&{44.83}/44.79&{45.76}/45.72\\
			Nest-DGIL& 30.29/34.10&	39.62/34.56&\underline{43.11}/42.23&44.19/42.91&41.43/40.80&\underline{44.17}/43.70&	\underline{44.97}/44.91&\underline{45.85}/45.78 \\
			{DPIR} & {\underline{35.77}} &	{36.83} &	{37.13} &	{37.31} &{41.18} &	{42.02} &	{42.44} &	{42.82} \\
			\textbf{MVMS-RCN} & \textbf{40.81}&\textbf{42.62}&\textbf{43.98}&\textbf{45.64}& \textbf{43.35} &\textbf{44.38} &\textbf{45.13} &\textbf{46.11}\\
			\bottomrule[1.5pt]
	\end{tabular*}}
\end{table*}

% fig denoiser
\begin{figure*}
	\centerline{\includegraphics[width=2\columnwidth]{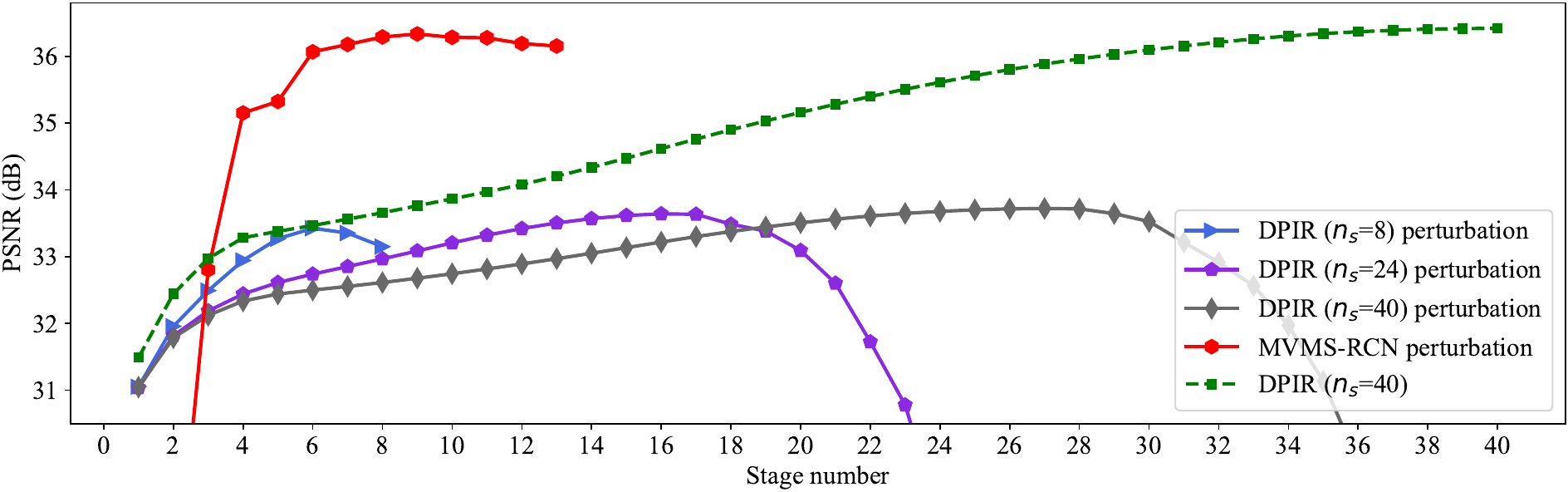}}
	\caption{{The convergence curve for the reconstruction of the perturbed projection data in a fan-beam projection with 32 views.}}
	\label{fig_denoiser}
\end{figure*}

\subsubsection{Qualitative evaluation}
Fig. \ref{fig_comparison_CT} shows the results of CT reconstruction using different methods in the fan-beam projection with 32 views. We can see that the FBP reconstruction has a lot of noise and artifacts. ISTA-Net, ISTA-Net+, FISTA-Net and Nest-DGIL can reduce some of the noise and streaking artifacts, but details are lost. RED-CNN, FBPConvNet and Uformer can remove streaking artifacts effectively, but some small structures may be smoothed out. PD-Net can remove streaking artifacts well and reconstruct high-quality CT images, but the details and texture information are not completely preserved. Although DPIR employs CG to maintain projection data consistency, it results in blurring of many fine details. Our results (l1)-(l4) have the best performance in eliminating noise artifacts while retaining intricate details (e.g., tiny blood vessels and bronchi).

Fig. \ref{fig_comparison_CT_parallel} illustrates a comparison of CT reconstructions using parallel-beam projection with 60 views.  It can be observed that the FBP-reconstructed CT images are of poor quality, with a lot of noise and artifacts. The other methods
are able to reduce the noise and remove most of the artifacts, but they fail to preserve the details and texture information.
Our methods (l1)-(l2) are superior to the other competitive reconstruction methods, as they refine the multi-view projection errors and apply multi-scale geometric correction, resulting in better artifact removal and detail preservation.

\subsection{Blind Reconstruction of Untrained Sparse-view Projections}
Table \ref{table_result_untrained} shows the quantitative comparison of blind reconstructions for untrained sparse-view projections. All compared methods, except for FBP, DRUNet, and DPIR, train separate models for each fan-beam projection task (with 16, 32, 64, 128, and 256 views) and parallel-beam projection tasks (having 30, 60, 90, 120, and 180 views). Conversely, the proposed framework utilizes a single model to handle five sparse-view CT reconstruction tasks. For methods that lack the capability to reconstruct CT images directly from untrained sparse-view projections, their PSNR results are evaluated using the two nearest trained sparse-view models.

Deep unfolding methods (PD-Net, ISTA-Net, ISTA-Net+ and Nest-DGIL) transform truncated unfolding optimization into an end-to-end trainable deep network. However, these methods typically require task-specific training, potentially reducing their effectiveness in practical applications. In contrast, our method removes the necessity for such specialized training, providing a more adaptable and easily implementable solution across various scenarios. This flexibility arises from our technique of refining projection errors from both projected and non-projected perspectives in both full-view and sparse-view environments, presenting a novel method to utilize available projection data. Furthermore, the CNN denoiser in DRUNet and DPIR lacks a comprehensive analysis and explanation. We convert the completion of details in the image domain into correcting errors of different scales related to the target image’s geometric errors across multi-scale and multi-resolution levels. By aligning the detailed mathematical framework with the network architecture, our approach ensures that all mathematical operations are systematically integrated into the network design. Consequently, the transition from mathematical theory to network design is more natural compared to existing deep unfolding methods.

Our framework outperforms the other methods for fan-beam projection, with an average improvement of 2.46 dB. Specifically, it improves at least 5.05 dB, 2.97 dB, 0.87 dB, and 0.97 dB for 24 views, 48 views, 96 views, and 192 views, respectively. For parallel-beam projection, our framework improves 1.28 dB, 0.21 dB, 0.16 dB, and 0.27 dB for 45 views, 75 views, 105 views, and 150 views, respectively, with an average improvement of 0.48 dB. It is noteworthy that our method can use the single multi-sparse-view model to produce more satisfactory results with any untrained sparse-view projections than the other compared methods.

\subsection{Plug-and-Play Sparse-view CT Reconstruction}
Due to the shared multi-scale geometric correction module $\mathcal{D}$, the proposed MVMS-RCN can serve as a plug-and-play denoiser \cite{Zhang2022a} for sparse-view CT reconstruction. To bridge the gap between simulated and real medical data, we introduced a perturbation of $±1\%$ to the distances between the rotation center and both the source and the detector, enhancing the realism of the medical CT reconstruction process. The plug-and-play sparse-view CT reconstruction results for the perturbed data, using fan-beam projection with 32 views, are shown in Fig. \ref{fig_denoiser}.

When encountering perturbed projection data, DPIR's reconstruction performance unfortunately suffers from an undesirable decline due to over-fitting. However, MVMS-RCN demonstrates exceptional stability, with the reconstruction results continually improving with each iteration. Remarkably, MVMS-RCN achieves optimal performance after the 9th iteration, and the following iterations maintain consistent results without significant degradation, effectively minimizing the risk of over-fitting. This highlights the potential of MVMS-RCN as a flexible plug-and-play solution designed for sparse-view CT reconstruction. Unlike DPIR, which frequently demands manual parameter tuning, MVMS-RCN removes this requirement, significantly enhancing reproducibility and efficiency. The ability of MVMS-RCN to easily adapt and succeed in challenging scenarios involving perturbed data highlights its robustness and adaptability for practical applications, marking it as a promising advancement in the field of CT reconstruction.

% fig unsupervised
\begin{figure}
	\centerline{\includegraphics[width=\columnwidth]{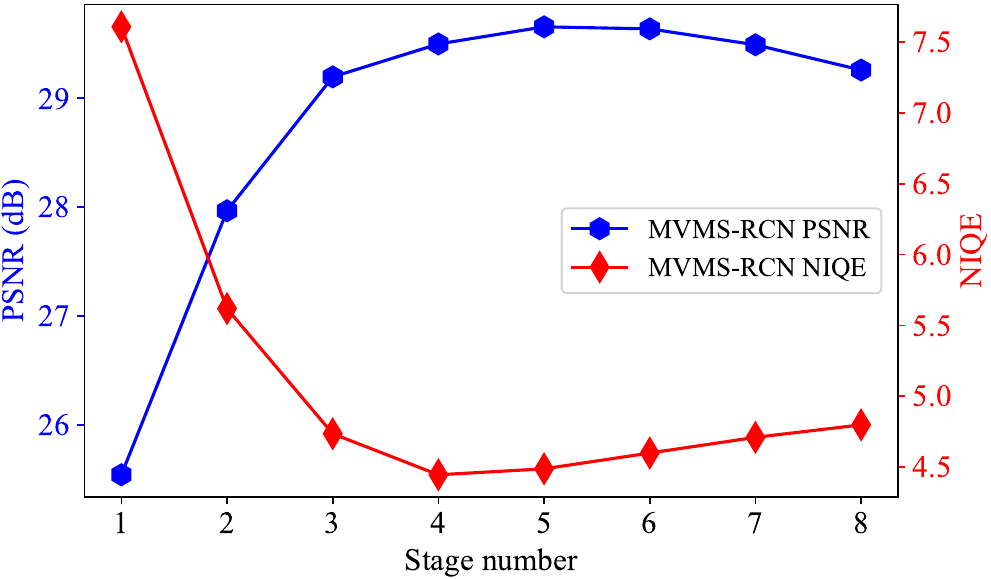}}
	\caption{The convergence curve for the reconstruction of the clinically rebinned projection data in a fan-beam projection with 36 views ($\times$32).}
	\label{fig_rebin}
\end{figure}

\subsection{Unsupervised Fine-tuning for Clinical Reconstruction}
In our clinical study concentrated on unsupervised fine-tuning, we utilize the same validation and test raw projection data from the 'Low Dose CT Image and Projection Data' as detailed in Section \ref{data_detail}. To improve computational efficiency, we adapt the different scanning configuration \cite{Wagner2023} for unsupervised training and evaluation, as detailed in Table 1 of \cite{Wu2024}, with a total of 1152 views and 736 detectors. The projection data for the validation and test patients are rebinned into 18, 36, 72, 144, and 288 views, with a voxel size of 0.7 in each slice. We select 362 slices from the validation patient and 385 slices from the test patient for use in unsupervised training and evaluation.

To adapt the pre-trained unified plug-and-play framework for a new scanning configuration, we minimize a loss function that operates without ground truth, defined by 
\begin{equation*}
\begin{split}
\mathcal{L}_{\text{total}}(\boldsymbol{\Theta})=&\frac{1}{d} \left\|\boldsymbol{P}_s^T\boldsymbol{P}_s\boldsymbol{x}_{n_s}-\boldsymbol{P}_s^T\boldsymbol{y}_s\right\|_{1}\\&+\gamma \left(1-{\text{SSIM}}\left(\boldsymbol{P}_s^T\boldsymbol{P}_s\boldsymbol{x}_{n_s},\boldsymbol{P}_s^T\boldsymbol{y}_s\right)\right)
\label{eq14}
\end{split}
\end{equation*}
through unsupervised fine-tuning across 30 epochs using the validation dataset. We also use the no-reference image quality assessment metric known as the Natural Image Quality Evaluator (NIQE) \cite{Mittal2013} to assess plug-and-play clinically rebinned reconstruction, Fig.\ref{fig_rebin} shows the convergence curve of the unsupervised sparse-view CT reconstruction derived from 36 views. The NIQE convergence trend closely aligns with that of PSNR, highlighting our method's potential for unsupervised applications with clinical data.

\section{Conclusion}
In this study, we propose a novel interpretable dual-domain framework that can flexibly deal with multi-sparse-view CT imaging in a single model. The proposed framework integrates the theoretical advantages of model-based methods and the superior reconstruction performance of DL-based methods, providing the generalizability of DL. It gives us a new way to explore the potential projection information, offers a new perspective on designing an interpretable network, and serves as a flexible model for multi-sparse-view CT reconstruction. Experiments show that the proposed framework outperforms other state-of-the-art methods in terms of qualitative and quantitative evaluaions. {In future work, our aim is to enhance the unified framework to simultaneously reconstruct fan-beam and parallel-beam projection data, and to extend its applicability to 3D reconstruction, limited-angle CT, and real medical CT image reconstruction problems. Our ultimate goal is to develop a more comprehensive and adaptable reconstruction framework.}

%% Loading bibliography style file
\bibliographystyle{IEEEtran}
\bibliography{mvms-ref-tci-2024}

% Generated by IEEEtran.bst, version: 1.14 (2015/08/26)
\begin{thebibliography}{10}
\providecommand{\url}[1]{#1}
\csname url@samestyle\endcsname
\providecommand{\newblock}{\relax}
\providecommand{\bibinfo}[2]{#2}
\providecommand{\BIBentrySTDinterwordspacing}{\spaceskip=0pt\relax}
\providecommand{\BIBentryALTinterwordstretchfactor}{4}
\providecommand{\BIBentryALTinterwordspacing}{\spaceskip=\fontdimen2\font plus
\BIBentryALTinterwordstretchfactor\fontdimen3\font minus
  \fontdimen4\font\relax}
\providecommand{\BIBforeignlanguage}[2]{{%
\expandafter\ifx\csname l@#1\endcsname\relax
\typeout{** WARNING: IEEEtran.bst: No hyphenation pattern has been}%
\typeout{** loaded for the language `#1'. Using the pattern for}%
\typeout{** the default language instead.}%
\else
\language=\csname l@#1\endcsname
\fi
#2}}
\providecommand{\BIBdecl}{\relax}
\BIBdecl

\bibitem{Natterer2001}
F.~Natterer, \emph{The Mathematics of Computerized Tomography}.\hskip 1em plus
  0.5em minus 0.4em\relax {SIAM}, 2001.

\bibitem{Pan2009}
X.~Pan, E.~Y. Sidky, and M.~Vannier, ``Why do commercial {CT} scanners still
  employ traditional, filtered back-projection for image reconstruction?''
  \emph{Inverse Problems}, vol.~25, no.~12, p. 123009, 2009.

\bibitem{Abbas2013}
S.~Abbas, T.~Lee, S.~Shin, R.~Lee, and S.~Cho, ``Effects of sparse sampling
  schemes on image quality in low-dose {CT},'' \emph{Medical Physics}, vol.~40,
  no.~11, p. 111915, 2013.

\bibitem{Kim2015}
K.~Kim, J.~C. Ye, W.~Worstell, J.~Ouyang, Y.~Rakvongthai, G.~E. Fakhri, and
  Q.~Li, ``Sparse-view spectral {CT} reconstruction using spectral patch-based
  low-rank penalty,'' \emph{{IEEE} Transactions on Medical Imaging}, vol.~34,
  no.~3, pp. 748--760, 2015.

\bibitem{Louis1986}
A.~K. Louis, ``Incomplete data problems in x-ray computerized tomography,''
  \emph{Numerische Mathematik}, vol.~48, no.~3, pp. 251--262, 1986.

\bibitem{LaRoque2008}
S.~J. LaRoque, E.~Y. Sidky, and X.~Pan, ``Accurate image reconstruction from
  few-view and limited-angle data in diffraction tomography,'' \emph{Journal of
  the Optical Society of America A}, vol.~25, no.~7, p. 1772, 2008.

\bibitem{Yang2010a}
J.~Yang, H.~Yu, M.~Jiang, and G.~Wang, ``High-order total variation
  minimization for interior tomography,'' \emph{Inverse Problems}, vol.~26,
  no.~3, p. 035013, 2010.

\bibitem{Zeng2016}
D.~Zeng, J.~Huang, H.~Zhang, Z.~Bian, S.~Niu, Z.~Zhang, Q.~Feng, W.~Chen, and
  J.~Ma, ``Spectral {CT} image restoration via an average image-induced
  nonlocal means filter,'' \emph{{IEEE} Transactions on Biomedical
  Engineering}, vol.~63, no.~5, pp. 1044--1057, 2016.

\bibitem{Rantala2006}
M.~Rantala, S.~Vanska, S.~Jarvenpaa, M.~Kalke, M.~Lassas, J.~Moberg, and
  S.~Siltanen, ``Wavelet-based reconstruction for limited-angle x-ray
  tomography,'' \emph{{IEEE} Transactions on Medical Imaging}, vol.~25, no.~2,
  pp. 210--217, 2006.

\bibitem{Dong2012}
B.~Dong, J.~Li, and Z.~Shen, ``X-ray {CT} image reconstruction via wavelet
  frame based regularization and radon domain inpainting,'' \emph{Journal of
  Scientific Computing}, vol.~54, no. 2-3, pp. 333--349, 2013.

\bibitem{Bao2019}
P.~Bao, H.~Sun, Z.~Wang, Y.~Zhang, W.~Xia, K.~Yang, W.~Chen, M.~Chen, Y.~Xi,
  S.~Niu, J.~Zhou, and H.~Zhang, ``Convolutional sparse coding for compressed
  sensing {CT} reconstruction,'' \emph{{IEEE} Transactions on Medical Imaging},
  vol.~38, no.~11, pp. 2607--2619, 2019.

\bibitem{Lu2023}
S.~Lu, B.~Yang, Y.~Xiao, S.~Liu, M.~Liu, L.~Yin, and W.~Zheng, ``Iterative
  reconstruction of low-dose {CT} based on differential sparse,''
  \emph{Biomedical Signal Processing and Control}, vol.~79, p. 104204, 2023.

\bibitem{Gao2011}
H.~Gao, J.~Cai, Z.~Shen, and H.~Zhao, ``Robust principal component
  analysis-based four-dimensional computed tomography,'' \emph{Physics in
  Medicine and Biology}, vol.~56, no.~11, pp. 3181--3198, 2011.

\bibitem{Semerci2014}
O.~Semerci, N.~Hao, M.~E. Kilmer, and E.~L. Miller, ``Tensor-based formulation
  and nuclear norm regularization for multienergy computed tomography,''
  \emph{{IEEE} Transactions on Image Processing}, vol.~23, no.~4, pp.
  1678--1693, 2014.

\bibitem{Zhao2012}
B.~Zhao, H.~Ding, Y.~Lu, G.~Wang, J.~Zhao, and S.~Molloi, ``Dual-dictionary
  learning-based iterative image reconstruction for spectral computed
  tomography application,'' \emph{Physics in Medicine and Biology}, vol.~57,
  no.~24, pp. 8217--8229, 2012.

\bibitem{Zhang2017b}
Y.~Zhang, X.~Mou, G.~Wang, and H.~Yu, ``Tensor-based dictionary learning for
  spectral {CT} reconstruction,'' \emph{{IEEE} Transactions on Medical
  Imaging}, vol.~36, no.~1, pp. 142--154, 2017.

\bibitem{Dong2018}
Y.~Dong, P.~C. Hansen, and H.~M. Kjer, ``Joint {CT} reconstruction and
  segmentation with discriminative dictionary learning,'' \emph{{IEEE}
  Transactions on Computational Imaging}, vol.~4, no.~4, pp. 528--536, 2018.

\bibitem{Li2022a}
S.~Li, Q.~Li, R.~Li, W.~Wu, J.~Zhao, Y.~Qiang, and Y.~Tian, ``An adaptive
  self-guided wavelet convolutional neural network with compound loss for
  low-dose {CT} denoising,'' \emph{Biomedical Signal Processing and Control},
  vol.~75, p. 103543, 2022.

\bibitem{Chen2017}
H.~Chen, Y.~Zhang, M.~K. Kalra, F.~Lin, Y.~Chen, P.~Liao, J.~Zhou, and G.~Wang,
  ``Low-dose {CT} with a residual encoder-decoder convolutional neural
  network,'' \emph{{IEEE} Transactions on Medical Imaging}, vol.~36, no.~12,
  pp. 2524--2535, 2017.

\bibitem{Jin2017}
K.~H. Jin, M.~T. McCann, E.~Froustey, and M.~Unser, ``Deep convolutional neural
  network for inverse problems in imaging,'' \emph{{IEEE} Transactions on Image
  Processing}, vol.~26, no.~9, pp. 4509--4522, 2017.

\bibitem{He2016}
K.~He, X.~Zhang, S.~Ren, and J.~Sun, ``Deep residual learning for image
  recognition,'' in \emph{{IEEE} Conference on Computer Vision and Pattern
  Recognition ({CVPR})}, 2016, pp. 770--778.

\bibitem{Zhang2018c}
Z.~Zhang, X.~Liang, X.~Dong, Y.~Xie, and G.~Cao, ``A sparse-view {CT}
  reconstruction method based on combination of {DenseNet} and deconvolution,''
  \emph{{IEEE} Transactions on Medical Imaging}, vol.~37, no.~6, pp.
  1407--1417, 2018.

\bibitem{Huang2017}
G.~Huang, Z.~Liu, L.~V.~D. Maaten, and K.~Q. Weinberger, ``Densely connected
  convolutional networks,'' in \emph{{IEEE} Conference on Computer Vision and
  Pattern Recognition ({CVPR})}, 2017, pp. 2261--2269.

\bibitem{Ronneberger2015}
O.~Ronneberger, P.~Fischer, and T.~Brox, ``U-net: Convolutional networks for
  biomedical image segmentation,'' in \emph{Medical Image Computing and
  Computer Assisted Intervention {\textendash} {MICCAI}}, 2015, pp. 234--241.

\bibitem{Han2018}
Y.~Han and J.~C. Ye, ``Framing {U}-net via deep convolutional framelets:
  Application to sparse-view {CT},'' \emph{{IEEE} Transactions on Medical
  Imaging}, vol.~37, no.~6, pp. 1418--1429, 2018.

\bibitem{Huang2022}
Z.~Huang, J.~Zhang, Y.~Zhang, and H.~Shan, ``{DU}-{GAN}: Generative adversarial
  networks with dual-domain u-net-based discriminators for low-dose {CT}
  denoising,'' \emph{{IEEE} Transactions on Instrumentation and Measurement},
  vol.~71, pp. 1--12, 2022.

\bibitem{Wang2023}
D.~Wang, F.~Fan, Z.~Wu, R.~Liu, F.~Wang, and H.~Yu, ``{CTformer}:
  convolution-free token2token dilated vision transformer for low-dose {CT}
  denoising,'' \emph{Physics in Medicine and Biology}, vol.~68, no.~6, p.
  065012, 2023.

\bibitem{Zhou2022}
B.~Zhou, X.~Chen, S.~K. Zhou, J.~S. Duncan, and C.~Liu, ``Dudodr-net:
  Dual-domain data consistent recurrent network for simultaneous sparse view
  and metal artifact reduction in computed tomography,'' \emph{Medical Image
  Analysis}, vol.~75, p. 102289, 2022.

\bibitem{CheslereanBoghiu2023}
T.~Cheslerean-Boghiu, F.~C. Hofmann, M.~Schulthei, F.~Pfeiffer, D.~Pfeiffer,
  and T.~Lasser, ``Wnet: A data-driven dual-domain denoising model for
  sparse-view computed tomography with a trainable reconstruction layer,''
  \emph{IEEE Transactions on Computational Imaging}, vol.~9, pp. 120--132,
  2023.

\bibitem{Zhang2021a}
Y.~Zhang, T.~Lv, R.~Ge, Q.~Zhao, D.~Hu, L.~Zhang, J.~Liu, Y.~Zhang, Q.~Liu,
  W.~Zhao, and Y.~Chen, ``{CD}-net: Comprehensive domain network with spectral
  complementary for {DECT} sparse-view reconstruction,'' \emph{{IEEE}
  Transactions on Computational Imaging}, vol.~7, pp. 436--447, 2021.

\bibitem{Wu2021}
W.~Wu, D.~Hu, C.~Niu, H.~Yu, V.~Vardhanabhuti, and G.~Wang, ``{DRONE}:
  Dual-domain residual-based optimization network for sparse-view {CT}
  reconstruction,'' \emph{{IEEE} Transactions on Medical Imaging}, vol.~40,
  no.~11, pp. 3002--3014, 2021.

\bibitem{Chen2018}
H.~Chen, Y.~Zhang, Y.~Chen, J.~Zhang, W.~Zhang, H.~Sun, Y.~Lv, P.~Liao,
  J.~Zhou, and G.~Wang, ``{LEARN}: Learned experts' assessment-based
  reconstruction network for sparse-data {CT},'' \emph{{IEEE} Transactions on
  Medical Imaging}, vol.~37, no.~6, pp. 1333--1347, 2018.

\bibitem{Adler2018}
J.~Adler and O.~Oktem, ``Learned primal-dual reconstruction,'' \emph{{IEEE}
  Transactions on Medical Imaging}, vol.~37, no.~6, pp. 1322--1332, 2018.

\bibitem{Zhang2018}
J.~Zhang and B.~Ghanem, ``{ISTA}-{N}et: Interpretable optimization-inspired
  deep network for image compressive sensing,'' in \emph{{IEEE}/{CVF}
  Conference on Computer Vision and Pattern Recognition ({CVPR})}, 2018, pp.
  1828--1837.

\bibitem{Daubechies2004}
I.~Daubechies, M.~Defrise, and C.~D. Mol, ``An iterative thresholding algorithm
  for linear inverse problems with a sparsity constraint,''
  \emph{Communications on Pure and Applied Mathematics}, vol.~57, no.~11, pp.
  1413--1457, 2004.

\bibitem{Elad2006}
M.~Elad, ``Why simple shrinkage is still relevant for redundant
  representations?'' \emph{{IEEE} Transactions on Information Theory}, vol.~52,
  no.~12, pp. 5559--5569, 2006.

\bibitem{Xiang2020}
J.~Xiang, Y.~Dong, and Y.~Yang, ``{FISTA}-net: Learning a fast iterative
  shrinkage thresholding network for inverse problems in imaging,''
  \emph{{IEEE} Transactions on Medical Imaging}, vol.~40, no.~5, pp.
  1329--1339, 2021.

\bibitem{Beck2009}
A.~Beck and M.~Teboulle, ``A fast iterative shrinkage-thresholding algorithm
  for linear inverse problems,'' \emph{{SIAM} Journal on Imaging Sciences},
  vol.~2, no.~1, pp. 183--202, 2009.

\bibitem{Zhang2021}
Z.~Zhang, Y.~Liu, J.~Liu, F.~Wen, and C.~Zhu, ``{AMP}-{N}et: Denoising-based
  deep unfolding for compressive image sensing,'' \emph{{IEEE} Transactions on
  Image Processing}, vol.~30, pp. 1487--1500, 2021.

\bibitem{Donoho2009}
D.~L. Donoho, A.~Maleki, and A.~Montanari, ``Message-passing algorithms for
  compressed sensing,'' in \emph{Proceedings of the National Academy of
  Sciences}, vol. 106, no.~45, 2009, pp. 18\,914--18\,919.

\bibitem{Fan2023a}
X.~Fan, Y.~Yang, K.~Chen, Y.~Feng, and J.~Zhang, ``Nest-{DGIL}:
  Nesterov-optimized deep geometric incremental learning for cs image
  reconstruction,'' \emph{IEEE Transactions on Computational Imaging}, vol.~9,
  pp. 819--833, 2023.

\bibitem{Dutta2024}
S.~Dutta, A.~Basarab, B.~Georgeot, and D.~Kouamé, ``{DIVA}: Deep unfolded
  network from quantum interactive patches for image restoration,''
  \emph{Pattern Recognition}, vol. 155, p. 110676, 2024.

\bibitem{Dutta2021}
S.~Dutta, A.~Basarab, B.~Georgeot, and D.~Kouame, ``Image denoising inspired by
  quantum many-body physics,'' in \emph{IEEE International Conference on Image
  Processing (ICIP)}, 2021, pp. 1619--1623.

\bibitem{Dutta2022}
S.~Dutta, A.~Basarab, B.~Georgeot, and D.~Kouamé, ``A novel image denoising
  algorithm using concepts of quantum many-body theory,'' \emph{Signal
  Processing}, vol. 201, p. 108690, 2022.

\bibitem{Zhang2022a}
K.~Zhang, Y.~Li, W.~Zuo, L.~Zhang, L.~Van~Gool, and R.~Timofte, ``Plug-and-play
  image restoration with deep denoiser prior,'' \emph{IEEE Transactions on
  Pattern Analysis and Machine Intelligence}, vol.~44, no.~10, pp. 6360--6376,
  2022.

\bibitem{Metzler2017}
M.~Christopher, M.~Ali, and B.~Richard, ``Learned d-amp: A principled cnn-based
  compressive image recovery algorithm,'' in \emph{Advances in Neural
  Information Processing Systems}, 2017.

\bibitem{Borgerding2017}
M.~Borgerding, P.~Schniter, and S.~Rangan, ``{AMP}-inspired deep networks for
  sparse linear inverse problems,'' \emph{{IEEE} Transactions on Signal
  Processing}, vol.~65, no.~16, pp. 4293--4308, 2017.

\bibitem{Aggarwal2019}
H.~K. Aggarwal, M.~P. Mani, and M.~Jacob, ``{MoDL}: Model-based deep learning
  architecture for inverse problems,'' \emph{{IEEE} Transactions on Medical
  Imaging}, vol.~38, no.~2, pp. 394--405, 2019.

\bibitem{Yang2020}
Y.~Yang, J.~Sun, H.~Li, and Z.~Xu, ``{ADMM}-{CSNet}: A deep learning approach
  for image compressive sensing,'' \emph{{IEEE} Transactions on Pattern
  Analysis and Machine Intelligence}, vol.~42, no.~3, pp. 521--538, 2020.

\bibitem{Zhang2021b}
H.~Zhang, B.~Liu, H.~Yu, and B.~Dong, ``{MetaInv}-net: Meta inversion network
  for sparse view {CT} image reconstruction,'' \emph{{IEEE} Transactions on
  Medical Imaging}, vol.~40, no.~2, pp. 621--634, 2021.

\bibitem{Fan2021}
X.~Fan, Y.~Yang, and J.~Zhang, ``Deep geometric distillation network for
  compressive sensing {MRI},'' in \emph{{IEEE} {EMBS} International Conference
  on Biomedical and Health Informatics ({BHI})}, 2021.

\bibitem{Xia2021}
W.~Xia, Z.~Lu, Y.~Huang, Z.~Shi, Y.~Liu, H.~Chen, Y.~Chen, J.~Zhou, and
  Y.~Zhang, ``{MAGIC}: Manifold and graph integrative convolutional network for
  low-dose {CT} reconstruction,'' \emph{{IEEE} Transactions on Medical
  Imaging}, vol.~40, no.~12, pp. 3459--3472, 2021.

\bibitem{Fan2023}
X.~Fan, Y.~Yang, K.~Chen, J.~Zhang, and K.~Dong, ``An interpretable {MRI}
  reconstruction network with two-grid-cycle correction and geometric prior
  distillation,'' \emph{Biomedical Signal Processing and Control}, vol.~84, p.
  104821, 2023.

\bibitem{McCormick1987}
S.~F. McCormick, ``Multigrid methods,'' \emph{{SIAM}}, 1987.

\bibitem{Xu1992}
J.~Xu, ``A new class of iterative methods for nonselfadjoint or indefinite
  problems,'' \emph{{SIAM} Journal on Numerical Analysis}, vol.~29, no.~2, pp.
  303--319, 1992.

\bibitem{Xu1996}
{J. Xu}, ``Two-grid discretization techniques for linear and nonlinear
  {PDEs},'' \emph{{SIAM} Journal on Numerical Analysis}, vol.~33, no.~5, pp.
  1759--1777, 1996.

\bibitem{Anselone1974}
P.~M. Anselone and J.~W. Lee, ``Spectral properties of integral operators with
  nonnegative kernels,'' \emph{Linear Algebra and its Applications}, vol.~9,
  pp. 67--87, 1974.

\bibitem{Song2021}
J.~Song, B.~Chen, and J.~Zhang, ``Memory-augmented deep unfolding network for
  compressive sensing,'' in \emph{Proceedings of the 29th {ACM} International
  Conference on Multimedia}, 2021, pp. 4249--4258.

\bibitem{KJiang2018}
K.~Jiang, Z.~Wang, P.~Yi, J.~Jiang, J.~Xiao, and Y.~Yao, ``Deep distillation
  recursive network for remote sensing imagery super-resolution,'' \emph{Remote
  Sensing}, vol.~10, no.~11, p. 1700, 2018.

\bibitem{He2015}
K.~He, X.~Zhang, S.~Ren, and J.~Sun, ``Delving deep into rectifiers: Surpassing
  human-level performance on {ImageNet} classification,'' in \emph{{IEEE}
  International Conference on Computer Vision ({ICCV})}, 2015, pp. 1026--1034.

\bibitem{Moen2020}
T.~R. Moen, B.~Chen, D.~R. Holmes, X.~Duan, Z.~Yu, L.~Yu, S.~Leng, J.~G.
  Fletcher, and C.~H. McCollough, ``Low-dose {CT} image and projection
  dataset,'' \emph{Medical Physics}, vol.~48, no.~2, pp. 902--911, 2020.

\bibitem{Ronchetti2020}
M.~Ronchetti, ``Torchradon: Fast differentiable routines for computed
  tomography,'' \emph{arXiv}, 2020.

\bibitem{Kingma2014}
K.~Diederik and B.~Jimmy, ``Adam: A method for stochastic optimization,'' in
  \emph{Proceedings of ICLR}, 2015.

\bibitem{Beck2009a}
A.~Beck and M.~Teboulle, ``Fast gradient-based algorithms for constrained total
  variation image denoising and deblurring problems,'' \emph{{IEEE}
  Transactions on Image Processing}, vol.~18, no.~11, pp. 2419--2434, 2009.

\bibitem{Wang2022}
Z.~Wang, X.~Cun, J.~Bao, W.~Zhou, J.~Liu, and H.~Li, ``Uformer: A general
  u-shaped transformer for image restoration,'' in \emph{{IEEE}/{CVF}
  Conference on Computer Vision and Pattern Recognition ({CVPR})}, 2022, pp.
  17\,662--17\,672.

\bibitem{Wagner2023}
F.~Wagner, M.~Thies, L.~Pfaff, O.~Aust, S.~Pechmann, D.~Weidner, N.~Maul,
  M.~Rohleder, M.~Gu, J.~Utz, F.~Denzinger, and A.~Maier, ``On the benefit of
  dual-domain denoising in a self-supervised low-dose ct setting,'' in
  \emph{IEEE 20th International Symposium on Biomedical Imaging (ISBI)},
  vol.~12, 2023, pp. 1--5.

\bibitem{Wu2024}
J.~Wu, X.~Jiang, L.~Zhong, W.~Zheng, X.~Li, J.~Lin, and Z.~Li, ``Linear
  diffusion noise boosted deep image prior for unsupervised sparse-view ct
  reconstruction,'' \emph{Physics in Medicine and Biology}, vol.~69, no.~16, p.
  165029, 2024.

\bibitem{Mittal2013}
A.~Mittal, R.~Soundararajan, and A.~C. Bovik, ``Making a “completely blind”
  image quality analyzer,'' \emph{IEEE Signal Processing Letters}, vol.~20,
  no.~3, pp. 209--212, 2013.

\end{thebibliography}

\end{document}